\documentclass[a4paper,11pt]{article}
\usepackage{aaskaiid}
\usepackage{titlesec}
\usepackage{comment}
\usepackage{orcidlink}
\usepackage{gensymb}

\title{Exploring the Influence of the Jet-environment Interactions on the Observed Asymmetry of Extragalactic Radio Sources with SKAO}
\ShortTitle{Exploring Jet-Environmet Interation in Radio Sources With SKAO}

\author[1]{Sabyasachi Pal\orcidlink{0000-0003-2325-8509}}
\ShortName{Pal et al.} 
\author[1]{Souvik Manik\orcidlink{0000-0002-6794-7405}}
\author[1]{Shobha Kumari\orcidlink{0000-0003-4213-9679}}
\author[2]{Chiranjib Konar\orcidlink{0000-0002-2530-3812}}

\affiliation[1]{Department of Pure and Applied Sciences, Midnapore City College, West Bengal, India}

\affiliation[2]{Department of Physics, Amity Institute of Applied Sciences, Amity University, Uttar Pradesh, Sector-125, Noida.  201313, U.P., India}
\emailAdd{sabya.pal@gmail.com}

\abstract{Several observational and theoretical studies have suggested that the observed arm-length asymmetries in extragalactic radio sources are primarily driven by interactions between radio jets and an inhomogeneous ambient medium, although orientation effects may also contribute to the observed asymmetry. However, the observational evidence supporting these interpretations comes from only a small sample of FR II radio sources, in which the brighter hotspots are typically found on the side of the shorter jet arm. We aim to investigate the interactions between powerful jets and their surrounding environments in radio sources, with a particular focus on how these interactions shape the morphology and asymmetry of the radio lobes. The unprecedented sensitivity and angular resolution of the Square Kilometre Array Observatory (SKAO) will facilitate the discovery and detailed characterization of asymmetric radio sources across a wide range of physical scales and redshifts. Using a combination of 3D magnetohydrodynamic simulations and observational data from the SKAO, one can explore the influence of clumpy interstellar and intergalactic media on jet propagation and the resulting asymmetries in radio sources at various redshifts. The study will analyze how environmental factors, such as density and turbulence, decelerate jets, leading to observable asymmetries in smaller, higher-redshift sources. In this chapter, we review existing simulation and observational results on jet-environment interactions in radio galaxies and discuss how SKAO capabilities will further advance our understanding of AGN feedback and its role in shaping large-scale cosmic structure.
}
\begin{document}
\maketitle

\section{Introduction and Motivation}
Active galactic nuclei (AGN) are powered by accretion onto supermassive black holes (SMBHs) and are among the most energetic objects in the Universe. A fraction of AGN produce collimated, relativistic jets that can travel tens to hundreds of kiloparsecs, inflating large radio lobes and cavities. As these jets propagate through the interstellar medium (ISM), circumgalactic medium (CGM), and intracluster medium (ICM), they strongly influence both the appearance of radio emission and the thermal and dynamical state of the surrounding gas.

Once a jet is launched, the environment becomes a key driver of its evolution. At low redshift, edge-darkened Fanaroff--Riley \citep{Fanaroff74} Type I (FR-I) radio galaxies are usually found in rich clusters, while edge-brightened FR-II systems tend to reside in poorer environments \citep{Hill91}. FR-IIs generally have higher radio luminosities than FR-Is, although the two classes show considerable overlap \citep{best09, Mira17}. They also interact differently with their environments: FR-Is inject relatively gentle, continuous heating into cluster cores \citep[e.g.][]{Churazov01, Fabian12}, whereas FR-IIs produce powerful shocks that can influence gas and satellite galaxies over hundreds of kiloparsecs \citep[e.g.][]{Rawlings2004, Shabala2011}.

The Fanaroff--Riley (FR) classification is strongly influenced by jet--environment interactions on kiloparsec scales. If the jet is slowed by entrainment of ambient gas \citep{Bicknell1995} or by stellar winds \citep{Komissarov1994, Perucho2014}, it becomes unstable and transitions into an FR-I morphology. If entrainment is weak, the jet maintains collimation and produces an FR-II structure with hotspots at the terminal shocks. Other mechanisms, such as jet stalling in a rising pressure atmosphere \citep{Massaglia2016} or failure of an initially conical jet to collimate \citep{Alexander2006, Krause2012}, can also lead to eventual jet disruption and FR-I–like structures.

A particularly interesting and rare class of radio AGN is the hybrid morphology radio sources (HyMoRS), where one lobe exhibits FR-I morphology and the opposite lobe shows FR-II morphology \citep{gopal_krishna_2000, Gaw06, kapinska17, harwood20, kumari22, manik25}. Although extremely uncommon ($<1\%$; \citealt{Gaw06,manik25}), these sources are excellent laboratories for studying environmental effects, since the two jets originate from the same central engine and differences between the lobes primarily reflect the environments the jets encounter.

Radio galaxies often exhibit significant asymmetries between their two lobes in terms of arm length, flux density, spectral index, and polarisation. Typical arm-length ratios ($Q = l_{\rm long}/l_{\rm short}$) are in the range $Q \sim 1.1-1.5$, although extreme cases can exceed $Q > 2$. Similarly, flux density ratios between the two lobes ($R = S_{\rm bright}/S_{\rm faint}$) commonly lie in the range $R \sim 2-5$, and occasionally reach values greater than $\sim10$ \citep[e.g.][]{manik2025grq}. Polarisation asymmetries are also frequently observed, largely due to differential Faraday depolarisation along the two lines of sight through the magnetoionic medium. Such asymmetries provide important diagnostics of jet propagation and the interaction between radio jets and their surrounding environment. These asymmetries contain important clues about jet physics, environmental density gradients, interactions with neighbouring galaxies, and relativistic beaming effects. The observed asymmetry in radio galaxies can be attributed to two broader reasons: (i) jet-environment interaction and (ii) relativistic effect in jets, even when the jet and counter jets are intrinsically the same. The interaction between jet and environment can differ not only due to the density of the ambient medium, but also due to whether the ambient medium is a non-relativistic thermal plasma medium (for radio galaxies) or a relativistic non-thermal plasma medium (for inner jets due to episodic jet activity in episodic radio galaxies). Of course, some ambient medium could be a mixture of both types. However, progress in understanding these asymmetries has long been limited by the sensitivity and dynamic range of previous radio surveys, which often miss faint counter-jets and low surface-brightness lobes.

At the extreme end of the asymmetry distribution are single-lobe radio galaxies, in which one lobe is either undetected or exhibits a flux density at least an order of magnitude lower than that of the opposite lobe. Early examples of such highly asymmetric systems were reported by \citet{Harris1984}, while later studies identified remarkable cases such as CGCG~049-033, where a prominent jet and lobe are visible on only one side of the nucleus \citep{Bagchi2007}. These objects provide valuable constraints on extreme environmental asymmetries or propagation effects that can strongly influence jet evolution.

Direct evidence of jet–environment interaction is also observed in cases where radio jets collide with companion galaxies or dense gas clouds. One of the best-known examples is Minkowski's object, where the radio jet from the galaxy NGC~541 is believed to trigger star formation in a nearby gas-rich system \citep{Croft2006}. Another recent example is RAD-12, where interaction between a radio jet and a neighbouring galaxy produces clear morphological distortions and complex radio structures \citep{Hota2022}. A further RAD@home discovery, a bow-and-arrow radio galaxy tracing a $\sim$560~kpc bow-shock structure in a multi-halo environment, provides additional evidence of jet-driven shocks shaping radio morphology on group/cluster scales \citep{Hota2026}. Although fewer than a dozen such jet–galaxy interaction systems are currently known, they provide unique laboratories for studying positive AGN feedback and jet-induced star formation.

Radio jets and lobes can also interact with structures on larger scales, such as galaxy filaments. In some cases, the flow of intrafilament medium can bend or distort radio lobes. For example, \citet{Edwards2010} reported a radio galaxy whose lobes appear bent due to motion through a large-scale filament associated with the cluster Abell~1763. Recently, \citet{manik2025grq} presented a new sample of 53 giant radio quasars (GRQs) discovered from the TGSS survey and investigated their large-scale environments. They found that at least $\sim13\%$ of the sources reside within galaxy groups or clusters, with a comparable fraction located in the vicinity of cosmic filaments. Furthermore, GRQs at higher redshifts tend to exhibit enhanced jet asymmetry, suggesting a stronger role of environmental effects in shaping radio source morphology at early cosmic epochs. A striking direct example of such large-scale environmental shaping is seen in a $\sim$560~kpc bow-shock structure recently discovered in a multi-halo environment \citep{Hota2026}. In a more recent study, \citet{Mahato2025} performed a statistical analysis of giant radio galaxies (GRGs) embedded in cosmic web filaments, showing that jet--filament alignment, rather than mere proximity to filaments, plays a key role in regulating the maximum source size. In this picture, GRGs with jets oriented at large angles to the filament spine can expand into lower-density, void-facing regions and grow to Mpc scales, whereas those aligned along the filament are more confined. 

\begin{figure}
    \centering
    \includegraphics[scale=0.4]{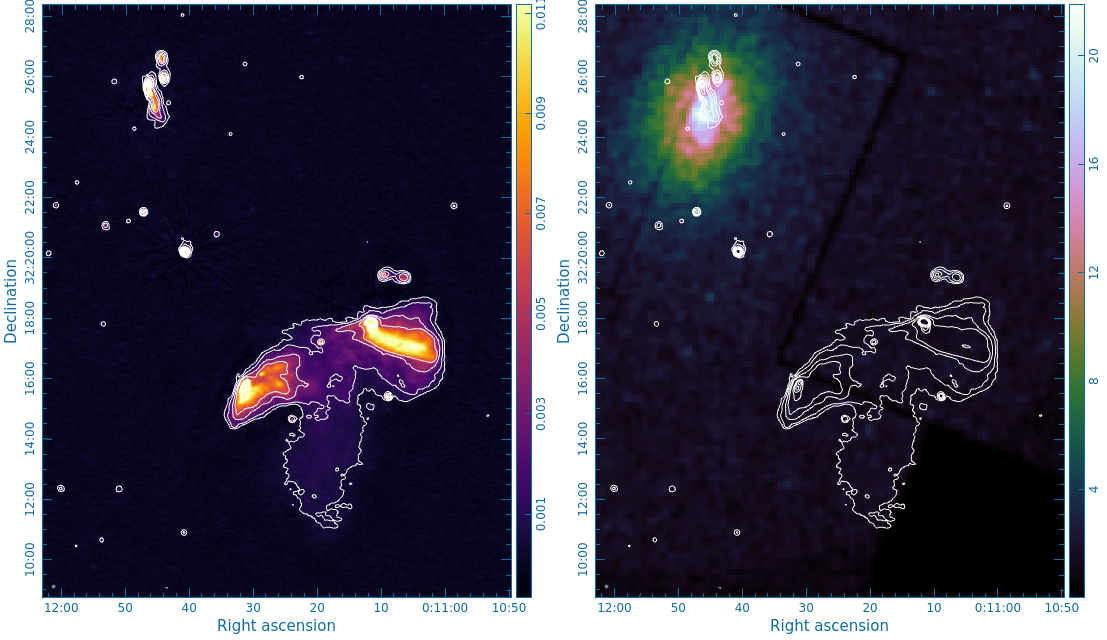}
    \caption{\textit{Left:} LoTSS 144~MHz radio continuum image of the peculiar giant radio galaxy J0011+3217, with white contours overlaid. The main source (Dec $\sim 32^\circ16'$--$32^\circ18'$) displays two bright primary lobes and a large one-sided diffuse secondary wing extending $\sim0.85$~Mpc to the south and west constituting $\sim$85\% of the primary lobe extent, unprecedented among known X-shaped radio galaxies. \textit{Right:} The same radio contours overlaid on the XMM-Newton smoothed X-ray map (0.5--2~keV). The X-ray emission peaks strongly toward the north, tracing the hot intracluster medium of the galaxy cluster Abell~7 ($z = 0.104$, $M_{500} = 3.71 \times 10^{14}\ M_\odot$), whose centre lies $\sim$1.2~Mpc from the radio source. The offset between the X-ray peak and the radio morphology, combined with the one-sided distortion of the secondary wing, is consistent with ram pressure exerted by the intracluster medium as J0011+3217 moves through the outskirts of Abell~7. This system illustrates how the large-scale thermal environment can impose dramatic, asymmetric morphological distortions on a megaparsec-scale radio source \citep{Kumari2024}.}
    \label{fig:J0011}
\end{figure}

The interplay between radio morphology and the thermal environment is strikingly illustrated by the peculiar giant radio galaxy J0011+3217 (Figure~\ref{fig:J0011}; \citealt{Kumari2024}). LoTSS 144~MHz imaging reveals misaligned primary lobes spanning $\sim$1~Mpc alongside a one-sided diffuse secondary wing extending $\sim$0.85~Mpc, a wing-to-lobe ratio of $\sim$85\%. The XMM-Newton X-ray map shows that the cluster Abell~7 dominates the thermal emission to the north, while the radio source itself sits in the cluster outskirts where X-ray surface brightness is low. The morphological asymmetry and southward distortion of the secondary wing are attributed to ram pressure from the ambient intracluster medium, providing a vivid, directly observed example of how large-scale thermal environment drives extreme radio asymmetry on megaparsec scales.

On scales of tens to hundreds of kiloparsecs, environmental structure predominantly governs lobe evolution rather than jet dynamics. Semi-analytic dynamical models \citep{Ka97, Blundell2000,turner15,Hardcastle18} and numerical simulations \citep{Hardcastle13, Hardcastle14} predict that lobe growth and luminosity evolution depend strongly on the ambient density profile; X-ray observations corroborate these predictions \citep{Arnaud10}. Compact radio AGN are more common in low-mass hosts and sparse environments \citep{Shabala18}, consistent with scenarios in which extended emission falls below present surface-brightness sensitivity limits \citep{Shabala17,TurnerEA18b}. Deep observations of the giant lobes of 3C 31 support this interpretation \citep{heesen18}.

Because the environment regulates lobe expansion and radiative evolution, estimates of jet kinetic power, AGN lifetime, and the total energy available for feedback are inherently environment dependent. This dependence is critical because AGN feedback plays a central role in shaping the bright end of the galaxy luminosity function \citep{silk98, croton6, Bower6}. Quantifying radio-source environments, for example via galaxy clustering, therefore provides a direct connection between galaxy-formation models and radio-lobe evolution. Semi-analytic frameworks that incorporate environmental information \citep{turner15} successfully reproduce many observed properties of radio galaxies. Furthermore, joint modelling of radio AGN and galaxy populations \citep{Shabala9, Raouf17} demonstrates that simultaneously matching both sets of observables places strong constraints on AGN feedback prescriptions.

The Square Kilometre Array Observatory (SKAO) will significantly advance our ability to study these processes. With its exceptional sensitivity, angular resolution, and dynamic range, SKA-Mid and SKA-Low will detect faint counter-lobes, relic plasma, restarted jets, and subtle morphological asymmetries that have been undetectable so far. The SKAO will enable detailed morphological, spectral, and polarization studies across a wide range of redshifts and environments, providing a comprehensive view of both young and evolved radio sources.

By enabling precise measurements of radio asymmetry and jet--environment interactions, the SKAO will provide powerful constraints on models of jet propagation, AGN feedback, and the co-evolution of SMBHs and their host galaxies. This chapter explores the physical origins of asymmetry, the diagnostic power of high-sensitivity observations, and the potential of combining SKAO data with multiwavelength surveys to advance our understanding of AGN feedback and its role in shaping large-scale cosmic structure.

\section{Background: Jet Asymmetry and Environmental Influence}

\subsection{Origin of Asymmetry in Radio Sources}

Extragalactic radio sources, particularly those associated with powerful radio-loud AGN, often display pronounced asymmetries in their large-scale structures. These asymmetries are observed in several key forms: differences in the apparent lengths of the two jets (\textit{arm-length asymmetry}), variations in the brightness or flux densities of the lobes (\textit{flux asymmetry}), and disparities in the polarization and spectral properties between the two sides. The studies of such asymmetries are not merely observational curiosities; they provide fundamental insights into the physical processes governing jet formation, propagation, and their interaction with the surrounding medium.

On parsec scales, relativistic beaming and Doppler boosting can account for a substantial portion of the observed brightness differences, as the approaching jet appears enhanced while the receding jet is deboosted. However, at kiloparsec and megaparsec scales, where jet speeds become sub-relativistic, orientation effects alone are insufficient to explain the observed asymmetries. Instead, environmental factors become dominant. Differences in the external medium, such as variations in gas density, pressure, temperature, magnetic field configuration and thermal/nonthermal nature, can influence jet propagation and confinement. A jet encountering a denser region of the interstellar or intracluster medium experiences greater deceleration and enhanced radiative losses, producing shorter, brighter, or more distorted lobes compared to its counterpart.

The observed arm-length ratio ($Q = l_{\mathrm{long}} / l_{\mathrm{short}}$) and flux ratio ($R = S_{\mathrm{bright}} / S_{\mathrm{faint}}$) are therefore crucial quantitative indicators of such effects \citep[e.g.][]{GK04,manik2025grq}. Statistical analyses of these parameters across radio galaxy samples have demonstrated correlations with environmental asymmetry indicators, suggesting that interaction with asymmetric gas distributions significantly contributes to the overall radio morphology.

\subsection{Jet--Environment Coupling}

The interaction between relativistic jets and their environments is inherently hydrodynamic and magnetohydrodynamic in nature. Depending upon whether there is significant entrainment of thermal matter into the jets or not, the jet will assume either FR-I or FR-II morphology. FR-I jets then propagate to inflated plumes without any hotspots at the jet head, and there will be no backflow of non-thermal plasma emanating from the jets. However, when an FR-II jet propagates through the ambient media, it drives two kinds of shocks, one of the strong bow shocks around the jet-head and a jet termination shock (sometimes called backward shock) at the jet head, which is the termination point. The non-thermal plasma of the jet naturally has to pass through the jet-termination shock before emanating from the jet-head and forming a cocoon of shocked relativistic plasma. 

The advance speed of the jet head, $v_{h}$, depends on both the jet’s kinetic power ($P_{j}$) and the external ambient medium density ($\rho_{a}$). If the ambient medium is a thermal medium of temperature $T\sim 10^{7}-10^{8}$K, then the rest mass density will play a role, and the momentum balance equation for the jet head can be written as \citep{Kaiser00}:

\begin{equation}
v_{h} = \sqrt{\frac{P_j}{c\,A_{h}\,\rho_a^{0}}}
\label{eq:a}
\end{equation}

where $A_{h}$ is the jet head cross-sectional area, and $~\rho_a^{0}$ is the rest mass density of the ambient medium, and $c$ is the speed of light.  However, this is a semi-relativistic equation and will work only when the jet-head is moving with a non-relativistic speed, which is almost always the case in single or double radio galaxies. Thus, a density contrast between the two sides of the host galaxy can naturally lead to unequal jet lengths and brightness. 

However, when the jet-heads of inner jets of episodic radio galaxies move with relativistic speed through non-thermal plasma of outer lobes, dumped in the previous episodes, this equation will have to be replaced by a formula derived by \citet{Konar13}, which is given by 

\begin{equation}
\beta_{h} = \frac{1}{\,1 + \sqrt{\frac{\beta_{j}\, c\, A_{h}\, w_{a}}{P_{j}}}}
\label{eq:b}
\end{equation}
where $\beta_{h}$ and $\beta_{j}$ are the hotspot speed and the jet bulk speed in units of c, $w_{a}$ is the relativistic enthalpy density of the ambient medium, and the other variables are already defined. This equation has to be used for the inner jets for double-double radio galaxies. In this case, not only density contrast can lead to asymmetry in jets, but also the relativistic beaming and deeming effect due to relativistic hotspot motion will give rise to observable structural asymmetry at the jet heads of the inner jets of double-double radio galaxies. 


Numerical simulations have shown that when a jet interacts with an inhomogeneous or clumpy environment, the resulting shocks and turbulence can produce complex, asymmetric morphologies. Dense clouds in the path of the jet can induce local deflections, generate secondary shocks, and trigger radiative cooling \citep{Gabkrause11, yatesjones21}. These processes not only modify the radio appearance of the source but also influence the energy transfer from the jet to the surrounding gas, an essential aspect of AGN feedback.

Observationally, polarization asymmetries provide additional evidence of environmental influence. The so-called \textit{Laing--Garrington effect} \citep{Laing1988, Garrington1988}, in which the lobe on the jet side appears less depolarized, is generally attributed to differences in the line-of-sight Faraday depth. In this scenario, the receding lobe is viewed through a longer magnetoionic path length in the surrounding medium, resulting in stronger depolarization.
These polarization gradients serve as probes of the magnetized environment around radio galaxies, linking asymmetry to the structure of the circumgalactic medium.

In addition to environmental effects, asymmetries in flux density and spectral index may also arise due to light travel time effects when the jets are inclined to the line of sight. Because the approaching and receding lobes are observed at slightly different evolutionary times, the lobes may appear to have different spectral ageing signatures and brightness levels. Consequently, apparent asymmetries may occur even if the jets are intrinsically symmetric.

\subsection{Implications for Galaxy and Cluster Evolution}

Jet–environment coupling has broader implications beyond morphology. The kinetic energy deposited by AGN jets heats the surrounding gas, regulates cooling flows in clusters, and suppresses star formation in massive galaxies. Environmental asymmetry modulates this feedback process: a jet encountering denser gas transfers energy more efficiently but advances more slowly, leading to asymmetric cavity structures and uneven heating. Understanding the statistical distribution of asymmetry parameters, therefore, provides indirect constraints on the efficiency and spatial distribution of AGN feedback.

In summary, the observed asymmetries in extragalactic radio sources arise from a complex interplay between intrinsic jet properties and external environmental conditions. Distinguishing between these contributions requires high-sensitivity, high-resolution observations combined with physical modelling and simulations. The SKAO, with its unparalleled imaging capabilities, is poised to deliver precisely the data needed to disentangle these effects, offering new insights into both the microphysics of jet propagation and the macroscopic impact of AGN on cosmic structure formation.

\subsection{Jet Propagation Through Thermal and Non-thermal Ambient Medium}
\citet{Konar13} reported structural asymmetries in the inner jet heads of several double-double radio galaxies, where the hotspot on one side appears compact and nearly spherical, whereas the hotspot on the opposite side exhibits a more elongated or conical morphology. In their Fig. 6, they presented some examples of the structural asymmetry in the observed images of inner jet heads and interpreted it to be due to relativistic jet-head motion, where Equation \ref{eq:b} will hold good for momentum balance at the jet head. As per their interpretation, the round hotspot is approaching us, and the conical jet-head side is receding from us. For inner jet-heads, the ambient medium is non-thermal plasma with little entrained thermal matter in it. Therefore, while the rest mass density is very low, the internal energy of the non-thermal medium contributes sufficient mass via $E = mc^2$, leading to a relativistic mass density given by $\rho_{a} = \frac{\epsilon + \rho_a^{0} c^{2}}{c^{2}}$. Consequently, the ambient medium becomes dense enough to create the hotspot at the inner jet-heads \citep{Konar13}.

\section{The Role of the SKAO in Studying Asymmetry}
Throughout this chapter, we adopt the current SKAO array assembly terminology. AA* refers to the initially funded deployment comprising 144 SKA-Mid dishes (80 SKA dishes together with the 64 MeerKAT dishes) and 307 SKA-Low stations, while AA4 denotes the full design baseline consisting of 197 SKA-Mid dishes and 512 SKA-Low stations \citep{Braun2024}. Unless otherwise stated, quoted sensitivities and angular resolutions refer to the AA4 design baseline.

\subsection{Advancing Beyond Current Capabilities}

Over the past few decades, radio interferometers such as the VLA, LOFAR, and MeerKAT have provided critical insights into the structure and dynamics of radio galaxies. These facilities have revealed a variety of asymmetric and hybrid morphology radio sources (HyMoRS), offering tantalizing evidence of the complex interplay between jets and their environments \citep{gopal_krishna_2000, harwood20, kumari22, manik25}. However, even the most sensitive current instruments face limitations in surface-brightness sensitivity, dynamic range, and polarization accuracy. As a result, faint counter-lobes, diffuse relic plasma, and low-level polarization structures often remain undetected, hindering our ability to fully characterize asymmetry and its physical origin.

Current state-of-the-art surveys each probe only part of the parameter space required for comprehensive studies of radio-source asymmetry. LoTSS provides excellent low-frequency sky coverage through its wide-area survey \citep{Shimwell2026}, while the LoTSS deep fields programme reaches $\mu$Jy-level sensitivities comparable to those achieved by MIGHTEE \citep{Shimwell2025, Hale2025}. Furthermore, imaging with the international LOFAR baselines delivers sub-arcsecond angular resolution at comparable sensitivities \citep{deJong2024}. However, these deep and high-resolution observations are currently limited to relatively small sky areas. MIGHTEE attains similar continuum sensitivities at GHz frequencies but over a more limited sky coverage. The SKAO will bridge this gap by combining sub-arcsecond angular resolution with $\mu$Jy~beam$^{-1}$ continuum sensitivity across large sky areas, enabling statistically robust investigations of radio-source asymmetry across a wide range of redshift, luminosity, and environment.


The SKAO, at its AA* and AA4 configurations, will overcome these limitations by delivering an unprecedented combination of sensitivity, angular resolution, and frequency coverage. SKA-Mid, operating primarily between 0.35 and 15.4 GHz, will achieve sub-arcsecond resolution and sub-$\mu$Jy sensitivity, enabling the detection of low-surface-brightness features and precise flux and polarization measurements. SKA-Low, covering frequencies below 350 MHz, will trace the oldest, steep-spectrum electron populations, revealing fossil emission and spectral-curvature asymmetries that encode the past dynamical history of jets. At low frequencies, SKA-Low AA4 is expected to achieve continuum sensitivities of $\sim5$--$10\,\mu\mathrm{Jy}\,\mathrm{beam}^{-1}$ at $\sim150\,\mathrm{MHz}$ in one-hour integrations \citep{Braun2024}, representing roughly an order-of-magnitude improvement over the median sensitivity of LoTSS DR3. This improvement will substantially increase the accessible volume for detecting fossil plasma, relic lobes, and remnant radio galaxies.

These capabilities will allow researchers to probe both compact and extended radio sources across a wide range of redshifts. For nearby galaxies, the SKAO will resolve fine-scale jet structures, identifying interaction zones with dense gas clouds. For high-redshift sources, where environmental effects are amplified by evolving intergalactic conditions, SKA-Mid will provide the first statistically robust measurements of asymmetry parameters, bridging the gap between individual case studies and population-level analyses.

\subsection{Quantifying Environmental Influence}

A major strength of the SKAO lies in its ability to link radio morphological asymmetry with independent tracers of the surrounding environment. By combining SKAO continuum and polarization maps with complementary datasets from next-generation optical and X-ray surveys, such as LSST, Euclid, and eROSITA, one can quantify local galaxy density, thermal gas distribution, and magnetic field structure around radio sources.

Multi-frequency polarimetric observations will enable high-precision \emph{Faraday rotation measure (RM)} mapping, providing a direct probe of the magnetoionic environment. Variations in RM and depolarization between the two lobes can be used to infer differences in electron density and magnetic field strength, thereby establishing whether one jet propagates through a denser or more turbulent region. The combination of SKA-Mid and SKA-Low polarimetry will, for the first time, allow continuous frequency coverage from tens of MHz to several GHz, yielding reliable rotation measure synthesis and detailed depolarization modelling.

Furthermore, environmental asymmetry can be statistically quantified through correlations between measured radio asymmetry parameters, arm-length ratio ($Q$), flux ratio ($R$), and depolarization ratio ($DP$), and environmental indicators such as galaxy overdensity, X-ray luminosity, or cluster richness. Such cross-correlations, possible only with the large samples expected from SKAO surveys, will directly test the long-standing hypothesis that jet asymmetries are primarily driven by external density gradients rather than intrinsic jet properties.

\subsection{Transforming Population Studies}

The SKAO has formally approved no Key Science Projects or large-area continuum surveys. Nevertheless, a plausible future science case is a wide-area SKA-Mid Band-2 continuum survey reaching sensitivities of order $\sigma \sim 2\,\mu$Jy\,beam$^{-1}$ rms over several thousand square degrees. Such a survey would combine the depth currently achieved only in targeted MeerKAT deep fields with sky coverage comparable to the largest existing radio surveys, enabling the construction of unprecedented samples of resolved radio galaxies suitable for asymmetry studies.

For comparison, LoTSS DR3 spans approximately 88\% of the northern sky at 144\,MHz with a median sensitivity of $\sim92\,\mu$Jy\,beam$^{-1}$ and catalogues nearly 13.7 million radio sources \citep{Shimwell2026}. MIGHTEE DR1 reaches sensitivities of $\sim1.3$--$2.7\,\mu$Jy\,beam$^{-1}$ at GHz frequencies but over only $\sim20\,\mathrm{deg}^{2}$ \citep{Hale2025}. The EMU survey is expected to detect approximately 40 million radio sources over most of the southern sky at a sensitivity of $\sim20\,\mu$Jy\,beam$^{-1}$ \citep{Norris2021}. A future SKA-Mid continuum survey combining MIGHTEE-like depth with LoTSS-like sky coverage would provide an unprecedented sample of well-resolved radio AGN, enabling population-scale studies of jet asymmetry, environmental influence, and AGN feedback across cosmic time.


In addition to large-scale surveys, targeted deep fields with SKA-Mid and SKA-Low will probe faint, high-redshift AGN and low-power radio galaxies, where asymmetry may be most sensitive to environmental variation. These observations will establish a crucial empirical foundation for theoretical models of jet–environment coupling and AGN feedback.

\subsection{Complementarity with Simulations and Theory}

The SKAO’s data products will not only enable observational advances but also provide essential constraints for numerical modelling. Three-dimensional hydrodynamic and magnetohydrodynamic (MHD) simulations can replicate jet propagation through non-uniform media, predicting the resulting distributions of arm-length and brightness asymmetries under varying conditions of ambient density, turbulence, and magnetic field geometry. Comparison between such simulations and SKAO observations will refine our understanding of jet dynamics, entrainment processes, and the microphysics of shock acceleration.

Through these synergies, the SKAO will transform asymmetry studies from a primarily qualitative domain into a quantitative, multi-dimensional science. The combined observational and theoretical framework will enable robust inference of jet power, environmental density contrasts, and the efficiency of energy transfer between AGN and their host environments.

\section{Methodology and Observational Approach}

\subsection{Defining and Measuring Asymmetry Parameters}

A rigorous study of radio-source asymmetry requires well-defined quantitative metrics that capture the geometric, radiative, and polarization properties of the two lobes. The arm-length ratio ($Q$), flux density ratio ($R$), and depolarization ratio ($DP$) are among the most widely used observables for quantifying large-scale asymmetries.

\begin{itemize}

\item \textbf{Arm-length ratio} ($Q = l_{\mathrm{long}} / l_{\mathrm{short}}$): This parameter quantifies the relative extension of the two lobes and serves as a measure of propagation asymmetry \citep[e.g.][]{GK04, manik2025grq}. Values of $Q > 1$ indicate that one jet has advanced farther into the surrounding medium, potentially due to differences in the external density or jet--medium interaction. In well-studied samples of FR-II radio galaxies, $Q$ typically lies in the range $1.1$--$1.5$ \citep[e.g.][]{Best95, GK04}, although extreme cases can reach $Q > 2$--$3$, occasionally exceeding $Q \sim 4$--$5$ in the most asymmetric sources \citep[e.g.][]{Bagchi2007, manik2025grq}. Giant radio sources tend to display somewhat larger asymmetries on average, consistent with propagation through more inhomogeneous large-scale environments \citep{manik2025grq}.

\item \textbf{Flux density ratio} ($R = S_{\mathrm{bright}} / S_{\mathrm{faint}}$): This ratio compares the integrated brightness of the two lobes \citep[e.g.][]{GK04,manik2025grq}. Deviations from unity may arise from Doppler boosting, adiabatic expansion losses, or environmental confinement affecting energy dissipation in each lobe. Typical values in FR-II radio galaxies lie in the range $R \sim 1.5$--$4$, with a median close to $R \sim 2$ \citep[e.g.][]{GK04}. In moderately asymmetric sources, $R$ reaches $5$--$10$, while in extreme cases -- including single-lobe or highly one-sided radio galaxies -- values of $R \gtrsim 20$ have been reported \citep{Bagchi2007}.

\item \textbf{Depolarization ratio} ($DP = p_1 / p_2$, where $p_i$ is the fractional polarization of lobe $i$): This parameter traces asymmetries in the magnetoionic environment along the two lines of sight, arising from differential Faraday depolarization between the approaching and receding lobes, the well-known Laing--Garrington effect \citep{Laing1988, Garrington1988}. The radio lobe on the same side as the approaching jet generally exhibits higher fractional polarization because its emission traverses a shorter column of the magnetoionic halo of the host galaxy, whereas the receding lobe's emission passes through a longer path through that halo, accumulating greater Faraday depolarization. The lobe on the same side as the approaching jet typically exhibits higher fractional polarization (i.e., less depolarization) because its line of sight traverses less of the magnetized halo. Observed values of $DP$ span a wide range; in moderately asymmetric sources $DP \sim 1.5$--$3$, but ratios exceeding $DP \sim 5$--$10$ are found in sources embedded in dense, magnetized cluster atmospheres \citep[e.g.][]{Garrington1991, GK04}. Since $DP$ is sensitive to both Faraday depth and magnetic turbulence along the line of sight, it provides an independent environmental diagnostic complementary to $Q$ and $R$.

\end{itemize}

With the sub-arcsecond imaging fidelity and $\mu\mathrm{Jy}\,\mathrm{beam}^{-1}$-level sensitivity of the SKA-Mid, these parameters can be measured for statistically meaningful samples across a wide range of redshifts. High dynamic range imaging will also enable the detection of diffuse, low-surface-brightness structures, ensuring that both the primary and counter lobes are robustly characterized.

\subsection{Spectral Index and Injection Index Asymmetry}

In addition to geometric and polarization asymmetries, differences in spectral properties between the two lobes can provide important diagnostics of particle acceleration and energy losses. A useful parameter in this context is the spectral index asymmetry,
\[
\Delta\alpha = \alpha_{\mathrm{jet}} - \alpha_{\mathrm{counterjet}},
\]
which quantifies the difference in spectral steepening between opposite lobes \citep{manik25}.

The spectral properties of radio lobes are closely linked to the efficiency of particle acceleration at the jet termination shocks. The strength of these shocks depends on the relativistic Mach number ($M$) of the jet fluid with respect to the hotspot frame rather than the host galaxy frame \citep{Konar13}. If the hotspot advances too rapidly, the effective Mach number may decrease; in the limit $M<1$, a strong termination shock may not form at the jet head. Since the efficiency of particle acceleration depends on the Mach number, differences in jet–environment interaction on the two sides may lead to variations in the injection spectral index.

Previous studies \citep{Konar06,Jamrozy07,Konar08,Nandi2010,Konar12,Konar13b, Patra2024} have generally constrained a single injection spectral index for both lobes of a radio galaxy due to limitations in sensitivity. The sensitivity and wide frequency coverage of the SKAO will enable this measurement to be performed separately for the two lobes, allowing a direct test of injection index asymmetry.

The SKAO will also allow detailed comparisons between the injection spectral indices of inner and outer lobes in double–double radio galaxies (DDRGs). For example, \citet{Konar13} reported similar injection spectral indices in the inner and outer doubles of several DDRGs and interpreted this as evidence for comparable jet powers in the two activity episodes. With the improved sensitivity and resolution of the SKAO, these results can be tested with much larger samples and with spatially resolved spectral analysis of hotspot regions.

Moreover, the detailed hotspot structures of both inner and outer jets can be studied with unprecedented sensitivity, providing new insights into jet dynamics, particle acceleration processes, and the microphysics of relativistic shocks.

\subsection{Theoretical Predictions for Asymmetry in FR-II Sources}
\label{sec:FRII_asymmetry_theory}

Dynamical models of classical FR-II radio sources predict explicit scalings between source linear size $D$, radio continuum luminosity $L_\nu$, and the properties of the ambient atmosphere \citep[e.g.][]{ShabalaGodfrey13,Ka97,WillottEA99}.  
For an atmosphere with a power-law density profile
\[
\rho(r)=\rho_0\left(\frac{r}{r_0}\right)^{-\beta},
\]
the analytic scalings (see Eqs.~1 and 4 of \citealt{ShabalaGodfrey13}) can be written as
\begin{equation}
D \;\propto\; \rho_0^{-a},\qquad a\equiv\frac{1}{5-\beta},
\label{eq:D_vs_rho}
\end{equation}
and
\begin{equation}
L_\nu \;\propto\; \rho_0^{(5+s)/12}\; D^{\,B},
\qquad
B \equiv 3 - \frac{4+\beta}{3}\left(\frac{5+s}{4}\right),
\label{eq:L_vs_rhoD}
\end{equation}
where \(s\) is the electron energy power-law index at injection (typical values \(s\approx2.0\text{--}2.5\); see \citealt{Ka97,TurnerEA18b}).

Substituting Equation~\ref{eq:D_vs_rho} into Equation~\ref{eq:L_vs_rhoD} gives a direct dependence of radio luminosity on ambient density,
\begin{align}
L_\nu &\propto \rho_0^{(5+s)/12}\,\left(\rho_0^{-a}\right)^{B}
\;=\; \rho_0^{\,p},\label{eq:L_vs_rho_p}\\
p &\equiv \frac{5+s}{12} - a\,B
\;=\; \frac{5+s}{12}
- \frac{1}{5-\beta}\left[3 - \frac{4+\beta}{3}\left(\frac{5+s}{4}\right)\right].
\label{eq:exponent_p}
\end{align}

These relations allow immediate predictions for asymmetry when the two sides of the host have different characteristic densities. Let \(r\equiv \rho_1/\rho_2\) be the density contrast between side~1 and side~2. Then

\begin{itemize}
	\item the predicted arm-length ratio is
	\begin{equation}
	Q \equiv \frac{D_1}{D_2} \;=\; r^{-a}
	\;=\; r^{-1/(5-\beta)},
	\label{eq:Q_vs_r}
	\end{equation}
	\item the predicted luminosity ratio is
	\begin{equation}
	R_L \equiv \frac{L_{1}}{L_{2}} \;=\; r^{\,p},
	\label{eq:RL_vs_r}
	\end{equation}
	where \(a\) and \(p\) are given by Equations~\ref{eq:D_vs_rho} and \ref{eq:exponent_p}, respectively.
\end{itemize}


Here, $s \approx 2$--$2.5$ denotes the typical slope of the power-law electron energy distribution at the hotspot, as indicated by standard dynamical and spectral-ageing models \citep[e.g.][]{Ka97, WillottEA99, TurnerEA18b}. Assuming the typical values of $\beta$, between 0 and 2, evaluating Equation~\ref{eq:exponent_p} for the commonly used parameter combinations yields the following exponents for the luminosity–density relation \(L_\nu\propto\rho_0^{p}\):
\[
\begin{array}{llc}
(\beta,s) & & p \\[2pt]\hline
(0,\,2.0) & & 0.4500\\
(0,\,2.5) & & 0.5250\\
(2,\,2.0) & & 0.7500\\
(2,\,2.5) & & 0.8750\\
\end{array}
\]

As a concrete illustration, for a modest density contrast \(r=2\) (i.e. one side twice as dense as the other) we obtain:
\[
\begin{array}{lcl}
\text{arm-length ratio } Q &=& r^{-1/(5-\beta)}\\[4pt]
\text{(for } r=2\text{):} & & Q(\beta=0)=2^{-0.20}\approx0.871,\quad Q(\beta=2)=2^{-1/3}\approx0.794,
\end{array}
\]
So the jet advancing into the denser medium is predicted to be noticeably shorter.

The corresponding luminosity ratios for \(r=2\) are
\[
R_L(r=2)=2^{p}\quad\Rightarrow\quad
\begin{cases}
(\beta=0,s=2.0): & R_L\approx 2^{0.45}\approx1.37,\\
(\beta=0,s=2.5): & R_L\approx 2^{0.525}\approx1.44,\\
(\beta=2,s=2.0): & R_L\approx 2^{0.75}\approx1.68,\\
(\beta=2,s=2.5): & R_L\approx 2^{0.875}\approx1.83.
\end{cases}
\]
These examples show that even modest density asymmetries ($r\sim 2$) produce measurable differences in both arm length and lobe luminosity, with the effect amplified for steeper atmosphere profiles (larger \(\beta\)) and larger injection indices \(s\).

\subsubsection{Caveats}

The analytic scalings above assume simplified conditions (self-similar expansion, no spatially varying electron ageing beyond the assumed injection slope, and a direct mapping between lobe pressure and emitted luminosity). In practice, additional effects, including spectral ageing, departures from equipartition, local re-acceleration, and magnetic-field inhomogeneities, introduce scatter and may modify the observed luminosity asymmetry. Nevertheless, Equations~\ref{eq:Q_vs_r} and \ref{eq:RL_vs_r} provide useful baseline expectations to compare with SKAO observations and with synthetic radio maps derived from MHD simulations.

The analytical relations above establish the baseline expectations for how asymmetric environments imprint measurable signatures on FR-II morphology. Density contrasts of only a factor of two can already produce arm-length ratios of $Q\sim1.2$--1.4 for realistic atmospheric slopes ($\beta=0$--2), while the corresponding luminosity asymmetries depend sensitively on the injection index $s$, with steeper electron spectra amplifying the environmental dependence. These theoretical curves therefore provide a physically grounded framework against which observational measurements of arm-length, flux density, spectral index, and depolarization asymmetries can be interpreted. Importantly, they also define the dynamic range and sensitivity required to detect such asymmetries in faint high-redshift sources. In this context, next-generation facilities such as the SKAO will enable population-level tests of these predictions, allowing the role of environmental gradients, jet composition, and plasma ageing to be disentangled with unprecedented precision.

\begin{figure}
  \centering
  \includegraphics[scale=0.5]{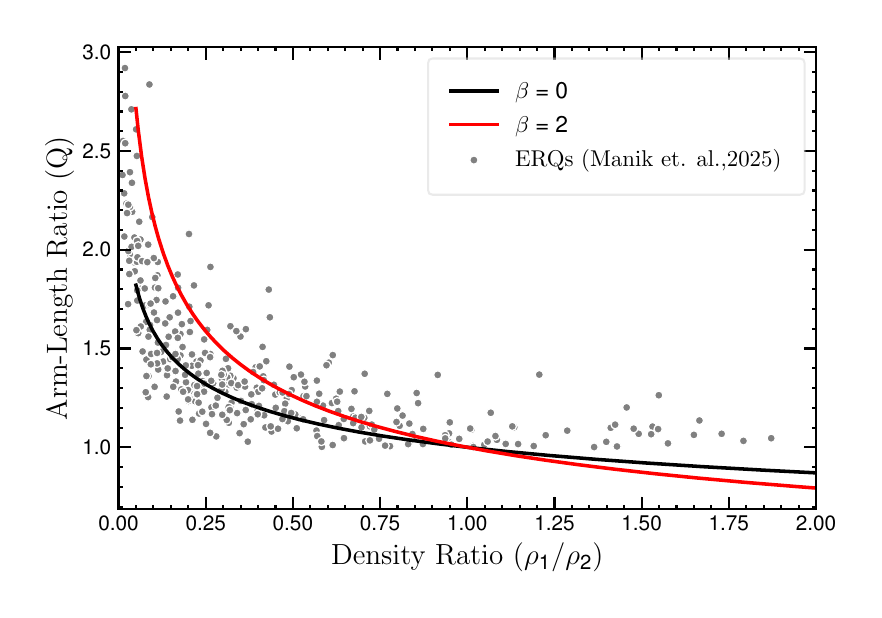}
    \includegraphics[scale=0.5]{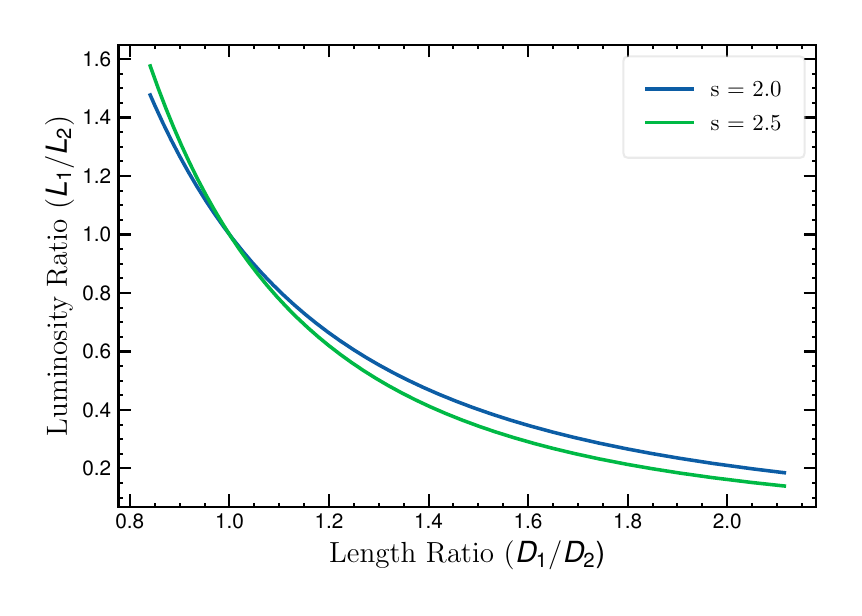}
  \caption{Predicted structural asymmetry for FR-II sources. \textit{Left:} Arm-length ratio $Q = D_1/D_2$ as a function of the ambient density contrast $r = \rho_1/\rho_2$. The curves drawn for $\beta = 0$ and $\beta = 2$. Grey points correspond to the estimated asymmetry properties of extended radio quasars from \citet{manik2025grq}. \textit{Right:} Luminosity ratio $R_L = L_1/L_2$ as a function of density contrast $r = \rho_1/\rho_2$, with curves shown for typical electron energy power-law index at injection $s \sim 2-2.5$.}
  \label{fig:asymmetry}
\end{figure}

\subsection{Survey Strategy and Source Selection}

A future wide-area continuum survey with SKA-Mid, if undertaken as part of a future SKAO science programme, would provide the primary dataset for statistical studies of radio-source asymmetry. For detailed analysis, a representative subsample of sources spanning a wide range of radio luminosities ($10^{23} \text{–} 10^{27}$ W Hz$^{-1}$), redshifts ($z \sim 0.1$–$2$), and morphological classes (FR I, FR II, and HyMoRS) will be selected.

Complementary deep observations in well-studied extragalactic fields (e.g., COSMOS, ELAIS-S1, and GOODS-South) will offer higher sensitivity to low-surface-brightness emission and provide access to extensive multiwavelength ancillary data. These deep fields will be critical for detecting faint counter-jets, relic lobes, and relic plasma that are essential for disentangling beaming from environmental effects.

Source selection will emphasize completeness and uniformity, ensuring that asymmetry statistics are not biased by orientation or selection thresholds. Automatic source extraction and morphological classification pipelines, supported by machine-learning algorithms trained on SKA Pathfinder data, will be used to identify double-lobed systems and measure structural parameters systematically.

\subsection{Cross-matching and Environmental Characterization}

Quantifying the environments of SKAO-detected radio galaxies is essential for interpreting jet asymmetry. Multiwavelength cross-matching will provide the host and environmental parameters needed to link asymmetry measures to underlying physical conditions.

Optical and infrared surveys (LSST, Euclid, JWST) will supply redshifts, stellar masses, and local galaxy-density estimates. X-ray missions (e.g.\ eROSITA, \textit{Chandra}) will map the hot intragroup and intracluster medium, yielding constraints on external density, pressure, and temperature. Spectroscopic surveys such as 4MOST \citep{4most} and DESI \citep{desi25} will add velocity-dispersion and dynamical information for clusters and groups.

Combining these datasets will allow the construction of a multi-parameter database linking radio asymmetry measures ($Q$, $R$, $DP$, $\Delta\alpha$) to physical environmental variables such as density contrast, magnetic field strength, and ambient pressure. Correlation analyses across this parameter space will test the hypothesis that environmental inhomogeneity is the dominant driver of large-scale radio asymmetry.

\subsection{Simulations and Model Comparisons}

Numerical simulations are essential for interpreting observed asymmetries in physical terms. Magnetohydrodynamic (MHD) and hydrodynamic simulations of jet propagation in stratified or clumpy media can reproduce a wide range of observed morphologies, including arm-length and brightness disparities \citep{Gabkrause11, yatesjones21}.

Model grids will be generated varying key parameters such as jet kinetic power, environmental density gradient, and magnetic field configuration. Synthetic radio maps derived from these simulations, convolved to SKAO-like resolution and noise levels, will be directly compared with observations. This comparison will constrain physical quantities such as jet advance speed, energy dissipation efficiency, and the role of magnetic fields in stabilizing or distorting jets.

The integration of observational data, statistical modelling, and high-resolution simulations will enable a physically grounded interpretation of the SKAO’s asymmetry measurements. In turn, these results will feed back into refined models of jet confinement, AGN feedback, and the influence of large-scale environment on radio-source evolution.

\section{Science Outcomes Enabled by SKAO}

The SKAO will open an unprecedented observational window for understanding the physical origin and cosmic evolution of radio source asymmetry. Its combination of high sensitivity, wide frequency coverage, and sub-arcsecond angular resolution will enable both population-level statistical studies and detailed investigations of individual systems. The ability to detect faint, extended, and low-surface-brightness features will significantly advance our understanding of jet-environment coupling, AGN duty cycles, and the feedback mechanisms that regulate galaxy evolution.

\subsection{Tracing Faint and Aged Radio Lobes}
One of the major advances expected from the SKAO is the ability to detect relic and remnant radio lobes in a statistically significant fraction of radio galaxies. Although such structures have already been identified in a limited number of sources in current low-frequency surveys, their true occurrence rate remains uncertain because of sensitivity limitations. These low-surface-brightness, steep-spectrum features trace previous episodes of AGN activity and therefore provide important constraints on AGN duty cycles.

SKA-Low, operating at frequencies below 350 MHz, will be particularly sensitive to this aged synchrotron emission, enabling the detection of faint relic plasma that remains invisible in most existing surveys. This will allow systematic studies of past jet activity and the processes that lead to morphological asymmetry. In particular, identifying asymmetric relic structures will help determine whether environmental confinement or intermittent jet activity is the dominant driver of the observed morphological differences.


\subsection{Quantifying Jet–Environment Interactions}

High dynamic range imaging from SKA-Mid will allow precise mapping of brightness and polarization asymmetries across radio lobes and jets. By combining these data with X-ray and optical surveys, it will be possible to correlate asymmetry parameters with host galaxy properties, intergalactic medium density, and local cluster conditions. This will enable quantitative tests of hydrodynamical models that describe jet deceleration and energy transfer. The SKAO will provide an unprecedented opportunity to investigate how environmental density gradients and magnetic fields influence jet propagation and radiative losses.

\subsection{Population Statistics and Evolution with Redshift}

With its wide-field survey capability, the SKAO will detect millions of radio sources across cosmic time, enabling statistical characterization of asymmetry as a function of redshift, radio power, and host morphology. Such large datasets will allow researchers to distinguish between intrinsic and extrinsic asymmetry drivers through population-level correlations. The combination of deep SKA-Mid and SKA-Low continuum surveys with optical and infrared data from LSST, Euclid, and JWST will help trace the co-evolution of AGN jets and their host galaxies, offering direct insights into how AGN feedback has evolved over billions of years.

\subsection{Polarization and Magnetic Field Diagnostics}

SKAO’s exceptional polarization sensitivity will enable detailed studies of Faraday rotation and depolarization asymmetry between the two lobes of radio galaxies. These diagnostics are powerful tools for probing magnetized plasma around jets and within the surrounding medium. By constructing rotation measure (RM) maps with high spatial precision, the SKAO will reveal how magnetic fields modulate jet stability and energy dissipation. This will further clarify the role of magnetic collimation and environmental magnetization in producing observed asymmetries.

\subsection{Discovery of Rare and Transitional Sources}

The SKAO’s sensitivity to both compact and diffuse emission will facilitate the discovery of rare hybrid and transitional systems, such as hybrid morphology radio sources (HyMoRS; \citealt{gopal_krishna_2000, Gaw06, kapinska17, harwood20, kumari22, manik25}) and double–double radio galaxies (DDRGs; \citealt{Konar13}), that offer unique insights into AGN intermittency and jet evolution. These sources provide natural laboratories for studying the interplay between intrinsic jet physics and asymmetric environmental conditions. Large samples of such systems will allow the development of new evolutionary models connecting jet reorientation, feedback episodes, and large-scale structure growth.

\subsection{Synergy with Multiwavelength and Numerical Studies}

Finally, the SKAO’s science outcomes will be amplified through synergy with next-generation observatories and simulations. Joint analyses with X-ray data from \textit{Athena}, optical/IR data from \textit{JWST} and \textit{Euclid}, and spectroscopic data from the 4MOST survey \citep{4most} will provide a multi-phase view of jet–environment interactions. In parallel, magnetohydrodynamical (MHD) simulations incorporating realistic density fields and feedback prescriptions will help interpret SKAO’s observations within a coherent theoretical framework. Together, these efforts will enable a comprehensive, multiwavelength picture of asymmetry formation and its role in the cosmic baryon cycle.






\section{Synergies and Staged Delivery}

The scientific potential of the SKAO in studying jet asymmetry and environmental influence will be realized progressively through its staged deployment and through coordinated synergy with other major astronomical facilities. The phased approach, beginning with SKA-Low and SKA-Mid and later progressing toward the AA4 design baseline, ensures that key science goals can be achieved incrementally while also guiding the refinement of data analysis techniques, calibration strategies, and theoretical modelling.

\subsection{Staged Delivery: From AA* to AA4}

The initial SKAO deployment (AA*) will already represent a transformative leap in radio astronomy.
\begin{itemize}
\item \textbf{SKA-Low} (50--350 MHz) will be essential for detecting faint, steep-spectrum, and aged synchrotron plasma associated with relic lobes and diffuse emission. Its sensitivity to low-frequency spectral curvature will help trace spectral ageing and identify remnant structures that encode the history of jet activity and asymmetry formation.
\item \textbf{SKA-Mid} (0.35--15.4 GHz) will provide sub-arcsecond resolution and high dynamic range imaging, enabling detailed mapping of brightness, polarization, and spectral index asymmetries in both compact and extended radio galaxies. Polarimetric observations will allow precise determination of Faraday rotation gradients and magnetic field configurations, yielding insights into the magneto-ionic environment and its asymmetrical effects on jet propagation.
\end{itemize}

Even during early operations (AA1--AA2), SKA pathfinder and precursor data (MeerKAT, ASKAP, uGMRT, LOFAR) will support preparatory science. These data can establish baseline asymmetry statistics and inform optimal survey strategies for SKA-Mid.

With Array Assembly 4 (AA4), full-sensitivity imaging will extend studies to sub-$\mu$Jy levels, revealing asymmetric emission in high-redshift AGN and faint populations of radio-quiet quasars.

A potential future expansion of the SKAO array beyond the AA4 design baseline could further extend both the frequency coverage and maximum baseline length, yielding substantial improvements in sensitivity and angular resolution. This will enable the detection of extremely faint counter-jets, low-luminosity sources at high redshifts ($z > 3$), and small-scale features such as jet knots or shock fronts. Combined with temporal monitoring, this will allow time-domain studies of evolving asymmetry, precession, and episodic activity across AGN lifecycles. However, the scope, technical specifications, and timeline of any such expansion remain to be defined.
 
\subsection{Synergies with Other Observatories}

SKAO science will be strengthened through coordinated multiwavelength observations, each providing complementary constraints on the AGN-environment system. X-ray facilities (\textit{Chandra}, \textit{XMM-Newton}, \textit{Athena}) will map the hot gas that shapes jet propagation. Optical and IR surveys (LSST, Euclid, JWST) will supply host-galaxy properties and merger histories. Millimetre/submillimetre observatories (ALMA, ngVLA) will trace cold gas and dust affected by jet-driven feedback. High-resolution MHD simulations and radiative-transfer modelling will further aid interpretation by linking SKAO observables to underlying jet–environment interactions.

\subsection{Progressive Science Returns}

The staged delivery of the SKAO ensures that critical science outcomes will be achieved early, even before full deployment. Pilot surveys conducted during AA* and AA4 commissioning will provide essential test cases for source finding, morphological classification, and polarization calibration pipelines. These early datasets will also enable cross-comparison with existing surveys, such as the LOFAR Two-Metre Sky Survey (LoTSS; \citealt{Shimwell17, Shimwell22}), the MeerKAT International GHz Tiered Extragalactic Exploration (MIGHTEE; \citealt{MIGHTEE, Hale2025}), and the Evolutionary Map of the Universe (EMU; \citealt{Norris2021}), thereby establishing continuity in the study of radio source asymmetry.

As the SKAO progresses from AA* to the full AA4 design baseline, the scope of scientific discovery will expand dramatically, from resolving individual jet-environment interactions to constructing a statistically robust evolutionary framework for AGN asymmetry across cosmic time.

\section{Future Enhancements and Prospects}
The scientific impact of the SKAO will continue to expand through advances in data-processing techniques, theoretical modelling, and coordinated observations with complementary facilities. The combination of increasingly sophisticated analysis methods and multiwavelength datasets will further enhance our ability to probe jet--environment interactions and the evolution of radio asymmetries across cosmic time.

\subsection{Enhanced Instrumentation and Capabilities}
The current AA4 design baseline for SKA-Mid covers frequencies between 0.35 and 15.4 GHz through Bands 1, 2, and 5 \citep{Bonaldi2024}. Investigations requiring observations at frequencies above 15 GHz will rely on complementary facilities such as ALMA and the ngVLA. Joint observations with these instruments will extend the spectral coverage available for studies of compact jet cores, hotspots, and particle acceleration processes. Within its planned frequency range, SKA-Mid will provide continuous multi-band polarimetric coverage capable of delivering high-fidelity Faraday rotation and depolarization measurements. Integration of SKA-Mid with existing VLBI networks, including the African VLBI Network and the European VLBI Network, will provide milliarcsecond-scale angular resolution for mapping compact jets, hotspots, and shock fronts, linking parsec- and kiloparsec-scale asymmetries \citep{ceglowski13}.

\subsection{Computational and Analytical Advances}
The vast data volumes produced by SKAO surveys necessitate innovative approaches to data handling and interpretation. Machine-learning and artificial intelligence (AI) frameworks, trained on precursor datasets from MeerKAT, ASKAP, and LOFAR, will enable the automatic identification and classification of asymmetric morphologies across the vast source populations expected from future SKAO surveys. These methods will help recognize rare phenomena such as hybrid morphology radio sources (HyMoRS), restarted jets, and precessing systems, cases where manual inspection would be infeasible.

Advanced statistical modelling, including Bayesian inference and neural simulation-based analyses, will allow direct comparison between observational distributions and theoretical predictions of jet–environment coupling. This will help constrain physical parameters such as jet power, density contrast, and magnetization, providing quantitative insights into asymmetry formation.



\subsection{Theoretical and Multi-Physics Modelling}
Coupled magnetohydrodynamic and radiative transfer simulations, combined with cosmic-ray transport and shock-cooling physics, will be crucial for interpreting SKAO observations \citep{Gabkrause11, yatesjones21}. These models will reproduce the morphological evolution of asymmetric jets in clumpy or stratified media, predicting observables such as arm-length ratio, spectral curvature, and polarization gradients. When compared with high-fidelity SKAO observations, such models will enable stringent tests of jet stability, entrainment, and feedback efficiency across a broad range of redshifts and environments.


\subsection{Broader Astrophysical Impact}

The insights gained from SKAO's study of jet asymmetry extend far beyond radio galaxy morphology. By tracing how AGN inject energy into the circumgalactic and intergalactic media, SKAO observations will illuminate fundamental processes governing galaxy evolution, quenching, and large-scale structure formation.
Moreover, asymmetric jet systems can serve as unique probes of anisotropic environments, cluster weather, and the role of cosmic magnetism in shaping extragalactic systems.

Ultimately, the SKAO, through its synergy with future facilities like \textit{Athena}, \textit{JWST}, \textit{Euclid}, and the \textit{ngVLA}, will play a central role in the next generation of multiwavelength astrophysical studies, enabling a comprehensive understanding of the interplay between black holes, jets, and their cosmic environments.

Compared with current facilities and surveys such as LoTSS, MIGHTEE, and EMU, the SKAO will combine significantly improved sensitivity, angular resolution, polarization fidelity, and survey speed within a single observatory. This capability will enable asymmetry studies to progress from investigations of individual sources and small samples to population-scale analyses involving hundreds of thousands of radio galaxies across a broad range of environments and cosmic epochs.



\section{Conclusions}

Jet asymmetry in extragalactic radio sources remains one of the most powerful probes of how relativistic outflows interact with their surrounding environments. The observed differences in arm length, flux density, spectral index, and polarization between opposite lobes encode vital information about both the intrinsic properties of the jets and the external media through which they propagate. Understanding the physical drivers of these asymmetries is therefore fundamental to disentangling the relative roles of jet dynamics, environmental structure, and feedback processes in galaxy evolution.

The unprecedented capabilities of the SKAO, in sensitivity, angular resolution, and broadband polarization fidelity, will revolutionize our capacity to explore these questions. SKA-Low and SKA-Mid will together enable comprehensive mapping of radio galaxies across a wide redshift range, capturing both active and relic emission components with unmatched detail. By correlating radio asymmetry parameters with environmental indicators derived from optical, infrared, and X-ray surveys, the SKAO will provide direct empirical tests of jet–medium coupling and feedback models.

Through a combination of deep imaging, polarization diagnostics, and large statistical samples, SKAO observations will finally allow the community to:
\begin{itemize}
\item Distinguish between intrinsic and environmental origins of asymmetry across diverse AGN populations;
\item Quantify the efficiency and spatial scale of jet-driven feedback within different galactic and cluster environments;
\item Trace the evolution of jet asymmetry and feedback strength from the nearby Universe to the epoch of early galaxy formation; and
\item Constrain the physical conditions, density, pressure, and magnetization, governing jet propagation in realistic cosmic settings.
\end{itemize}

By coupling these observations with next-generation simulations and machine-learning-based analysis pipelines, the SKAO will transform our theoretical and empirical understanding of radio galaxy asymmetry and its cosmological implications.
In doing so, the SKAO will not only address long-standing questions about jet physics but also establish radio asymmetry as a precise diagnostic tool for probing the structure and evolution of the cosmic environment from the local Universe to high redshift.

\bibliographystyle{abbrvnat-maxbibnames4}
\bibliography{chapter} 

@ARTICLE{gopal_krishna_2000,
       author = {{Gopal-Krishna} and {Wiita}, P.~J.},
        title = "{Extragalactic radio sources with hybrid morphology: implications for the Fanaroff-Riley dichotomy}",
      journal = {\aap},
         year = 2000,
        
       volume = {363},
        pages = {507-516},
          doi = {https://doi.org/10.48550/arXiv.astro-ph/0009441},
archivePrefix = {arXiv},
       eprint = {astro-ph/0009441},
 primaryClass = {astro-ph},
       adsurl = {https://ui.adsabs.harvard.edu/abs/2000A&A...363..507G}
}

@ARTICLE{harwood20,
       author = {{Harwood}, Jeremy J. and {Vernstrom}, Tessa and {Stroe}, Andra},
        title = "{Unveiling the cause of hybrid morphology radio sources (HyMoRS)}",
      journal = {\mnras},
         year = 2020,
        
       volume = {491},
       number = {1},
        pages = {803-822},
          doi = {https://doi.org/10.1093/mnras/stz3069},
archivePrefix = {arXiv},
       eprint = {1910.12857},
 primaryClass = {astro-ph.GA},
       adsurl = {https://ui.adsabs.harvard.edu/abs/2020MNRAS.491..803H}
}

@article{manik2025grq,
  title={Unveiling New Giant Radio Quasars from the TGSS Sky and their Large-scale Environment},
  author={Manik, Souvik and Bhukta, Netai and Pal, Sabyasachi and Mondal, Sushanta. K.},
  journal={\apjs},
  volume={281},
  pages={34},
  year={2025},
  doi = {https://doi.org/10.3847/1538-4365/ae0b52},
  publisher={American Astronomical Society}
}

@ARTICLE{Fanaroff74,
       author = {{Fanaroff}, B.~L. and {Riley}, J.~M.},
        title = "{The morphology of extragalactic radio sources of high and low luminosity}",
      journal = {\mnras},
         year = 1974,
        
       volume = {167},
        pages = {31P-36P},
          doi = {https://doi.org/10.1093/mnras/167.1.31P},
       adsurl = {https://ui.adsabs.harvard.edu/abs/1974MNRAS.167P..31F}
}

@ARTICLE{Hill91,
       author = {{Hill}, Gary J. and {Lilly}, Simon J.},
        title = "{A Change in the Cluster Environments of Radio Galaxies with Cosmic Epoch}",
      journal = {\apj},
         year = 1991,
        
       volume = {367},
        pages = {1},
          doi = {https://doi.org/10.1086/169597},
       adsurl = {https://ui.adsabs.harvard.edu/abs/1991ApJ...367....1H}
}

@ARTICLE{best09,
       author = {{Best}, P.~N.},
        title = "{Radio source populations: Results from SDSS}",
      journal = {Astronomische Nachrichten},
         year = 2009,
        
       volume = {330},
       number = {2},
        pages = {184-189},
          doi = {https://doi.org/10.1002/asna.200811152},
       adsurl = {https://ui.adsabs.harvard.edu/abs/2009AN....330..184B}
}

@ARTICLE{Mira17,
       author = {{Miraghaei}, H. and {Best}, P.~N.},
        title = "{The nuclear properties and extended morphologies of powerful radio galaxies: the roles of host galaxy and environment}",
      journal = {\mnras},
         year = 2017,
        
       volume = {466},
       number = {4},
        pages = {4346-4363},
          doi = {https://doi.org/10.1093/mnras/stx007},
archivePrefix = {arXiv},
       eprint = {1701.00919},
 primaryClass = {astro-ph.GA},
       adsurl = {https://ui.adsabs.harvard.edu/abs/2017MNRAS.466.4346M}
}

@ARTICLE{Churazov01,
       author = {{Churazov}, E. and {Br{\"u}ggen}, M. and {Kaiser}, C.~R. and {B{\"o}hringer}, H. and {Forman}, W.},
        title = "{Evolution of Buoyant Bubbles in M87}",
      journal = {\apj},
         year = 2001,
        
       volume = {554},
       number = {1},
        pages = {261-273},
          doi = {https://doi.org/10.1086/321357},
archivePrefix = {arXiv},
       eprint = {astro-ph/0008215},
 primaryClass = {astro-ph},
       adsurl = {https://ui.adsabs.harvard.edu/abs/2001ApJ...554..261C}
}

@ARTICLE{Fabian12,
       author = {{Fabian}, A.~C.},
        title = "{Observational Evidence of Active Galactic Nuclei Feedback}",
      journal = {\araa},
         year = 2012,
        
       volume = {50},
        pages = {455-489},
          doi = {https://doi.org/10.1146/annurev-astro-081811-125521},
archivePrefix = {arXiv},
       eprint = {1204.4114},
 primaryClass = {astro-ph.CO},
       adsurl = {https://ui.adsabs.harvard.edu/abs/2012ARA&A..50..455F}
}

@ARTICLE{Rawlings2004,
       author = {{Rawlings}, Steve and {Jarvis}, Matt J.},
        title = "{Evidence that powerful radio jets have a profound influence on the evolution of galaxies}",
      journal = {\mnras},
         year = 2004,
        
       volume = {355},
       number = {3},
        pages = {L9-L12},
          doi = {https://doi.org/10.1111/j.1365-2966.2004.08234.x},
archivePrefix = {arXiv},
       eprint = {astro-ph/0409687},
 primaryClass = {astro-ph},
       adsurl = {https://ui.adsabs.harvard.edu/abs/2004MNRAS.355L...9R}
}

@ARTICLE{Shabala2011,
       author = {{Shabala}, Stanislav S. and {Kaviraj}, Sugata and {Silk}, Joseph},
        title = "{Active galactic nucleus feedback drives the colour evolution of local galaxies}",
      journal = {\mnras},
         year = 2011,
        
       volume = {413},
       number = {4},
        pages = {2815-2826},
          doi = {https://doi.org/10.1111/j.1365-2966.2011.18353.x},
archivePrefix = {arXiv},
       eprint = {1102.3464},
 primaryClass = {astro-ph.CO},
       adsurl = {https://ui.adsabs.harvard.edu/abs/2011MNRAS.413.2815S}
}

@ARTICLE{Bicknell1995,
       author = {{Bicknell}, Geoffrey V.},
        title = "{Relativistic Jets and the Fanaroff-Riley Classification of Radio Galaxies}",
      journal = {\apjs},
         year = 1995,
        
       volume = {101},
        pages = {29},
          doi = {https://doi.org/10.1086/192232},
archivePrefix = {arXiv},
       eprint = {astro-ph/9406064},
 primaryClass = {astro-ph},
       adsurl = {https://ui.adsabs.harvard.edu/abs/1995ApJS..101...29B}
}

@ARTICLE{Komissarov1994,
       author = {{Komissarov}, S.~S.},
        title = "{Mass-Loaded Relativistic Jets}",
      journal = {\mnras},
         year = 1994,
        
       volume = {269},
        pages = {394},
          doi = {https://doi.org/10.1093/mnras/269.2.394},
       adsurl = {https://ui.adsabs.harvard.edu/abs/1994MNRAS.269..394K}
}

@ARTICLE{Perucho2014,
       author = {{Perucho}, M. and {Mart{\'\i}}, J.~M. and {Laing}, R.~A. and {Hardee}, P.~E.},
        title = "{On the deceleration of Fanaroff-Riley Class I jets: mass loading by stellar winds}",
      journal = {\mnras},
         year = 2014,
        
       volume = {441},
       number = {2},
        pages = {1488-1503},
          doi = {https://doi.org/10.1093/mnras/stu676},
archivePrefix = {arXiv},
       eprint = {1404.1209},
 primaryClass = {astro-ph.HE},
       adsurl = {https://ui.adsabs.harvard.edu/abs/2014MNRAS.441.1488P}
}

@ARTICLE{Massaglia2016,
       author = {{Massaglia}, S. and {Bodo}, G. and {Rossi}, P. and {Capetti}, S. and {Mignone}, A.},
        title = "{Making Faranoff-Riley I radio sources. I. Numerical hydrodynamic 3D simulations of low-power jets}",
      journal = {\aap},
         year = 2016,
        
       volume = {596},
          eid = {A12},
        pages = {A12},
          doi = {https://doi.org/10.1051/0004-6361/201629375},
archivePrefix = {arXiv},
       eprint = {1609.02497},
 primaryClass = {astro-ph.HE},
       adsurl = {https://ui.adsabs.harvard.edu/abs/2016A&A...596A..12M}
}

@ARTICLE{Alexander2006,
       author = {{Alexander}, P.},
        title = "{Models of young powerful radio sources}",
      journal = {\mnras},
         year = 2006,
        
       volume = {368},
       number = {3},
        pages = {1404-1410},
          doi = {https://doi.org/10.1111/j.1365-2966.2006.10225.x},
       adsurl = {https://ui.adsabs.harvard.edu/abs/2006MNRAS.368.1404A}
}

@ARTICLE{Krause2012,
       author = {{Krause}, Martin and {Alexander}, Paul and {Riley}, Julia and {Hopton}, Daniel},
        title = "{A new connection between the jet opening angle and the large-scale morphology of extragalactic radio sources}",
      journal = {\mnras},
         year = 2012,
        
       volume = {427},
       number = {4},
        pages = {3196-3208},
          doi = {https://doi.org/10.1111/j.1365-2966.2012.21645.x},
archivePrefix = {arXiv},
       eprint = {1206.1778},
 primaryClass = {astro-ph.CO},
       adsurl = {https://ui.adsabs.harvard.edu/abs/2012MNRAS.427.3196K}
}

@ARTICLE{Gaw06,
       author = {{Gawro{\'n}ski}, M.~P. and {Marecki}, A. and {Kunert-Bajraszewska}, M. and {Kus}, A.~J.},
        title = "{Hybrid morphology radio sources from the FIRST survey}",
      journal = {\aap},
         year = 2006,
        
       volume = {447},
       number = {1},
        pages = {63-70},
          doi = {https://doi.org/10.1051/0004-6361:20053996},
archivePrefix = {arXiv},
       eprint = {astro-ph/0509497},
 primaryClass = {astro-ph},
       adsurl = {https://ui.adsabs.harvard.edu/abs/2006A&A...447...63G}
}

@ARTICLE{kapinska17,
       author = {{Kapi{\'n}ska}, A.~D. and {Terentev}, I. and {Wong}, O.~I. and {Shabala}, S.~S. and {Andernach}, H. and {Rudnick}, L. and {Storer}, L. and {Banfield}, J.~K. and {Willett}, K.~W. and {de Gasperin}, F. and {Lintott}, C.~J. and {L{\'o}pez-S{\'a}nchez}, {\'A}. R. and {Middelberg}, E. and {Norris}, R.~P. and {Schawinski}, K. and {Seymour}, N. and {Simmons}, B.},
        title = "{Radio Galaxy Zoo: A Search for Hybrid Morphology Radio Galaxies}",
      journal = {\aj},
         year = 2017,
        
       volume = {154},
       number = {6},
          eid = {253},
        pages = {253},
          doi = {https://doi.org/10.3847/1538-3881/aa90b7},
archivePrefix = {arXiv},
       eprint = {1711.09611},
 primaryClass = {astro-ph.GA},
       adsurl = {https://ui.adsabs.harvard.edu/abs/2017AJ....154..253K}
}

@ARTICLE{kumari22,
       author = {{Kumari}, Shobha and {Pal}, Sabyasachi},
        title = "{Search for hybrid morphology radio galaxies from the FIRST survey at 1400 MHz}",
      journal = {\mnras},
         year = 2022,
        
       volume = {514},
       number = {3},
        pages = {4290-4299},
          doi = {https://doi.org/10.1093/mnras/stac1215},
archivePrefix = {arXiv},
       eprint = {2104.14469},
 primaryClass = {astro-ph.GA},
       adsurl = {https://ui.adsabs.harvard.edu/abs/2022MNRAS.514.4290K}
}

@ARTICLE{manik25,
       author = {{Manik}, Souvik and {Kumari}, Shobha and {Bhukta}, Netai and {Pal}, Sabyasachi and {Mondal}, Sushanta K.},
        title = "{Hybrid Morphology Radio Sources from the MeerKAT Absorption Line Survey (MALS): Radio, Mid-infrared, and Environmental Characteristics}",
      journal = {\apj},
         year = 2026,
       volume = {997},
       number = {2},
          eid = {157},
        pages = {157},
          doi = {https://doi.org/10.3847/1538-4357/ae1f85},
archivePrefix = {arXiv},
       eprint = {2506.20211},
 primaryClass = {astro-ph.GA},
       adsurl = {https://ui.adsabs.harvard.edu/abs/2026ApJ...997..157M}
}

@ARTICLE{Ka97,
       author = {{Kaiser}, Christian R. and {Alexander}, Paul},
        title = "{A self-similar model for extragalactic radio sources}",
      journal = {\mnras},
         year = 1997,
        
       volume = {286},
       number = {1},
        pages = {215-222},
          doi = {https://doi.org/10.1093/mnras/286.1.215},
       adsurl = {https://ui.adsabs.harvard.edu/abs/1997MNRAS.286..215K}
}

@ARTICLE{Blundell2000,
       author = {{Blundell}, Katherine M. and {Rawlings}, Steve},
        title = "{The Spectra and Energies of Classical Double Radio Lobes}",
      journal = {\aj},
         year = 2000,
        
       volume = {119},
       number = {3},
        pages = {1111-1122},
          doi = {https://doi.org/10.1086/301254},
archivePrefix = {arXiv},
       eprint = {astro-ph/0001327},
 primaryClass = {astro-ph},
       adsurl = {https://ui.adsabs.harvard.edu/abs/2000AJ....119.1111B}
}

@ARTICLE{turner15,
       author = {{Turner}, Ross J. and {Shabala}, Stanislav S.},
        title = "{Energetics and Lifetimes of Local Radio Active Galactic Nuclei}",
      journal = {\apj},
         year = 2015,
        
       volume = {806},
       number = {1},
          eid = {59},
        pages = {59},
          doi = {https://doi.org/10.1088/0004-637X/806/1/59},
archivePrefix = {arXiv},
       eprint = {1504.05204},
 primaryClass = {astro-ph.GA},
       adsurl = {https://ui.adsabs.harvard.edu/abs/2015ApJ...806...59T}
}

@ARTICLE{Hardcastle18,
       author = {{Hardcastle}, M.~J.},
        title = "{A simulation-based analytic model of radio galaxies}",
      journal = {\mnras},
         year = 2018,
        
       volume = {475},
       number = {2},
        pages = {2768-2786},
          doi = {https://doi.org/10.1093/mnras/stx3358},
archivePrefix = {arXiv},
       eprint = {1801.00667},
 primaryClass = {astro-ph.HE},
       adsurl = {https://ui.adsabs.harvard.edu/abs/2018MNRAS.475.2768H}
}

@ARTICLE{Hardcastle13,
       author = {{Hardcastle}, M.~J. and {Krause}, M.~G.~H.},
        title = "{Numerical modelling of the lobes of radio galaxies in cluster environments}",
      journal = {\mnras},
         year = 2013,
        
       volume = {430},
       number = {1},
        pages = {174-196},
          doi = {https://doi.org/10.1093/mnras/sts564},
archivePrefix = {arXiv},
       eprint = {1301.2531},
 primaryClass = {astro-ph.CO},
       adsurl = {https://ui.adsabs.harvard.edu/abs/2013MNRAS.430..174H}
}

@ARTICLE{Hardcastle14,
       author = {{Hardcastle}, M.~J. and {Krause}, M.~G.~H.},
        title = "{Numerical modelling of the lobes of radio galaxies in cluster environments - II. Magnetic field configuration and observability}",
      journal = {\mnras},
         year = 2014,
        
       volume = {443},
       number = {2},
        pages = {1482-1499},
          doi = {https://doi.org/10.1093/mnras/stu1229},
archivePrefix = {arXiv},
       eprint = {1406.5300},
 primaryClass = {astro-ph.HE},
       adsurl = {https://ui.adsabs.harvard.edu/abs/2014MNRAS.443.1482H}
}

@ARTICLE{Arnaud10,
       author = {{Arnaud}, M. and {Pratt}, G.~W. and {Piffaretti}, R. and {B{\"o}hringer}, H. and {Croston}, J.~H. and {Pointecouteau}, E.},
        title = "{The universal galaxy cluster pressure profile from a representative sample of nearby systems (REXCESS) and the Y$_{SZ}$ - M$_{500}$ relation}",
      journal = {\aap},
         year = 2010,
        
       volume = {517},
          eid = {A92},
        pages = {A92},
          doi = {https://doi.org/10.1051/0004-6361/200913416},
archivePrefix = {arXiv},
       eprint = {0910.1234},
 primaryClass = {astro-ph.CO},
       adsurl = {https://ui.adsabs.harvard.edu/abs/2010A&A...517A..92A}
}

@ARTICLE{Shabala18,
       author = {{Shabala}, Stanislav S.},
        title = "{The role of environment in the observed Fundamental Plane of radio active galactic nuclei}",
      journal = {\mnras},
         year = 2018,
        
       volume = {478},
       number = {4},
        pages = {5074-5080},
          doi = {https://doi.org/10.1093/mnras/sty1328},
archivePrefix = {arXiv},
       eprint = {1805.06600},
 primaryClass = {astro-ph.GA},
       adsurl = {https://ui.adsabs.harvard.edu/abs/2018MNRAS.478.5074S}
}

@ARTICLE{Shabala17,
       author = {{Shabala}, S.~S. and {Deller}, A. and {Kaviraj}, S. and {Middelberg}, E. and {Turner}, R.~J. and {Ting}, Y.~S. and {Allison}, J.~R. and {Davis}, T.~A.},
        title = "{Delayed triggering of radio active galactic nuclei in gas-rich minor mergers in the local Universe}",
      journal = {\mnras},
         year = 2017,
        
       volume = {464},
       number = {4},
        pages = {4706-4720},
          doi = {https://doi.org/10.1093/mnras/stw2536},
archivePrefix = {arXiv},
       eprint = {1608.04178},
 primaryClass = {astro-ph.GA},
       adsurl = {https://ui.adsabs.harvard.edu/abs/2017MNRAS.464.4706S}
}

@ARTICLE{TurnerEA18b,
       author = {{Turner}, Ross J. and {Shabala}, Stanislav S. and {Krause}, Martin G.~H.},
        title = "{RAiSE III: 3C radio AGN energetics and composition}",
      journal = {\mnras},
         year = 2018,
        
       volume = {474},
       number = {3},
        pages = {3361-3379},
          doi = {https://doi.org/10.1093/mnras/stx2947},
archivePrefix = {arXiv},
       eprint = {1711.04600},
 primaryClass = {astro-ph.GA},
       adsurl = {https://ui.adsabs.harvard.edu/abs/2018MNRAS.474.3361T}
}

@ARTICLE{heesen18,
       author = {{Heesen}, V. and {Croston}, J.~H. and {Morganti}, R. and {Hardcastle}, M.~J. and {Stewart}, A.~J. and {Best}, P.~N. and {Broderick}, J.~W. and {Br{\"u}ggen}, M. and {Brunetti}, G. and {Chy{\.Z}y}, K.~T. and {Harwood}, J.~J. and {Haverkorn}, M. and {Hess}, K.~M. and {Intema}, H.~T. and {Jamrozy}, M. and {Kunert-Bajraszewska}, M. and {McKean}, J.~P. and {Orr{\'u}}, E. and {R{\"o}ttgering}, H.~J.~A. and {Shimwell}, T.~W. and {Shulevski}, A. and {White}, G.~J. and {Wilcots}, E.~M. and {Williams}, W.~L.},
        title = "{LOFAR reveals the giant: a low-frequency radio continuum study of the outflow in the nearby FR I radio galaxy 3C 31}",
      journal = {\mnras},
         year = 2018,
        
       volume = {474},
       number = {4},
        pages = {5049-5067},
          doi = {https://doi.org/10.1093/mnras/stx2869},
archivePrefix = {arXiv},
       eprint = {1710.09746},
 primaryClass = {astro-ph.GA},
       adsurl = {https://ui.adsabs.harvard.edu/abs/2018MNRAS.474.5049H}
}

@ARTICLE{silk98,
       author = {{Silk}, Joseph and {Rees}, Martin J.},
        title = "{Quasars and galaxy formation}",
      journal = {\aap},
         year = 1998,
        
       volume = {331},
        pages = {L1-L4},
          doi = {https://doi.org/10.48550/arXiv.astro-ph/9801013},
archivePrefix = {arXiv},
       eprint = {astro-ph/9801013},
 primaryClass = {astro-ph},
       adsurl = {https://ui.adsabs.harvard.edu/abs/1998A&A...331L...1S}
}

@ARTICLE{Bower6,
       author = {{Bower}, R.~G. and {Benson}, A.~J. and {Malbon}, R. and {Helly}, J.~C. and {Frenk}, C.~S. and {Baugh}, C.~M. and {Cole}, S. and {Lacey}, C.~G.},
        title = "{Breaking the hierarchy of galaxy formation}",
      journal = {\mnras},
         year = 2006,
        
       volume = {370},
       number = {2},
        pages = {645-655},
          doi = {https://doi.org/10.1111/j.1365-2966.2006.10519.x},
archivePrefix = {arXiv},
       eprint = {astro-ph/0511338},
 primaryClass = {astro-ph},
       adsurl = {https://ui.adsabs.harvard.edu/abs/2006MNRAS.370..645B}
}

@ARTICLE{croton6,
       author = {{Croton}, Darren J. and {Springel}, Volker and {White}, Simon D.~M. and {De Lucia}, G. and {Frenk}, C.~S. and {Gao}, L. and {Jenkins}, A. and {Kauffmann}, G. and {Navarro}, J.~F. and {Yoshida}, N.},
        title = "{The many lives of active galactic nuclei: cooling flows, black holes and the luminosities and colours of galaxies}",
      journal = {\mnras},
         year = 2006,
        
       volume = {365},
       number = {1},
        pages = {11-28},
          doi = {https://doi.org/10.1111/j.1365-2966.2005.09675.x},
archivePrefix = {arXiv},
       eprint = {astro-ph/0508046},
 primaryClass = {astro-ph},
       adsurl = {https://ui.adsabs.harvard.edu/abs/2006MNRAS.365...11C}
}

@ARTICLE{Raouf17,
       author = {{Raouf}, Mojtaba and {Shabala}, Stanislav S. and {Croton}, Darren J. and {Khosroshahi}, Habib G. and {Bernyk}, Maksym},
        title = "{The many lives of active galactic nuclei-II: The formation and evolution of radio jets and their impact on galaxy evolution}",
      journal = {\mnras},
         year = 2017,
        month = oct,
       volume = {471},
       number = {1},
        pages = {658-670},
          doi = {https://doi.org/10.1093/mnras/stx1598},
archivePrefix = {arXiv},
       eprint = {1706.06595},
 primaryClass = {astro-ph.GA},
       adsurl = {https://ui.adsabs.harvard.edu/abs/2017MNRAS.471..658R}
}

@ARTICLE{Shabala9,
       author = {{Shabala}, Stanislav and {Alexander}, Paul},
        title = "{Radio Source Feedback in Galaxy Evolution}",
      journal = {\apj},
         year = 2009,
        
       volume = {699},
       number = {1},
        pages = {525-538},
          doi = {https://doi.org/10.1088/0004-637X/699/1/525},
archivePrefix = {arXiv},
       eprint = {0905.1578},
 primaryClass = {astro-ph.CO},
       adsurl = {https://ui.adsabs.harvard.edu/abs/2009ApJ...699..525S}
}

@ARTICLE{Kaiser00,
       author = {{Kaiser}, Christian R. and {Schoenmakers}, Arno P. and {R{\"o}ttgering}, Huub J.~A.},
        title = "{Radio galaxies with a `double-double' morphology - II. The evolution of double-double radio galaxies and implications for the alignment effect in FRII sources}",
      journal = {\mnras},
         year = 2000,
        
       volume = {315},
       number = {2},
        pages = {381-394},
          doi = {https://doi.org/10.1046/j.1365-8711.2000.03431.x},
archivePrefix = {arXiv},
       eprint = {astro-ph/9912142},
 primaryClass = {astro-ph},
       adsurl = {https://ui.adsabs.harvard.edu/abs/2000MNRAS.315..381K}
}

@ARTICLE{Konar13,
       author = {{Konar}, C. and {Hardcastle}, M.~J.},
        title = "{Particle acceleration and dynamics of double-double radio galaxies: theory versus observations}",
      journal = {\mnras},
         year = 2013,
        
       volume = {436},
       number = {2},
        pages = {1595-1614},
          doi = {https://doi.org/10.1093/mnras/stt1676},
archivePrefix = {arXiv},
       eprint = {1309.1401},
 primaryClass = {astro-ph.CO},
       adsurl = {https://ui.adsabs.harvard.edu/abs/2013MNRAS.436.1595K}
}

@ARTICLE{ShabalaGodfrey13,
       author = {{Shabala}, S.~S. and {Godfrey}, L.~E.~H.},
        title = "{Size Dependence of the Radio-luminosity-Mechanical-power Correlation in Radio Galaxies}",
      journal = {\apj},
         year = 2013,
        
       volume = {769},
       number = {2},
          eid = {129},
        pages = {129},
          doi = {https://doi.org/10.1088/0004-637X/769/2/129},
archivePrefix = {arXiv},
       eprint = {1304.6089},
 primaryClass = {astro-ph.CO},
       adsurl = {https://ui.adsabs.harvard.edu/abs/2013ApJ...769..129S}
}

@ARTICLE{WillottEA99,
       author = {{Willott}, Chris J. and {Rawlings}, Steve and {Blundell}, Katherine M. and {Lacy}, Mark},
        title = "{The emission line-radio correlation for radio sources using the 7C Redshift Survey}",
      journal = {\mnras},
         year = 1999,
        
       volume = {309},
       number = {4},
        pages = {1017-1033},
          doi = {https://doi.org/10.1046/j.1365-8711.1999.02907.x},
archivePrefix = {arXiv},
       eprint = {astro-ph/9905388},
 primaryClass = {astro-ph},
       adsurl = {https://ui.adsabs.harvard.edu/abs/1999MNRAS.309.1017W}
}

@ARTICLE{Konar06,
       author = {{Konar}, C. and {Saikia}, D.~J. and {Jamrozy}, M. and {Machalski}, J.},
        title = "{Spectral ageing analysis of the double-double radio galaxy J1453+3308}",
      journal = {\mnras},
         year = 2006,
        month = oct,
       volume = {372},
       number = {2},
        pages = {693-702},
          doi = {https://doi.org/10.1111/j.1365-2966.2006.10874.x},
archivePrefix = {arXiv},
       eprint = {astro-ph/0607660},
 primaryClass = {astro-ph},
       adsurl = {https://ui.adsabs.harvard.edu/abs/2006MNRAS.372..693K}
}

@ARTICLE{Jamrozy07,
       author = {{Jamrozy}, M. and {Konar}, C. and {Saikia}, D.~J. and {Stawarz}, {\L}. and {Mack}, K.-H. and {Siemiginowska}, A.},
        title = "{Intermittent jet activity in the radio galaxy 4C29.30?}",
      journal = {\mnras},
         year = 2007,
        
       volume = {378},
       number = {2},
        pages = {581-593},
          doi = {https://doi.org/10.1111/j.1365-2966.2007.11782.x},
archivePrefix = {arXiv},
       eprint = {astro-ph/0703723},
 primaryClass = {astro-ph},
       adsurl = {https://ui.adsabs.harvard.edu/abs/2007MNRAS.378..581J}
}

@ARTICLE{Konar08,
       author = {{Konar}, C. and {Jamrozy}, M. and {Saikia}, D.~J. and {Machalski}, J.},
        title = "{A multifrequency study of giant radio sources - I. Low-frequency Giant Metrewave Radio Telescope observations of selected sources}",
      journal = {\mnras},
         year = 2008,
        
       volume = {383},
       number = {2},
        pages = {525-538},
          doi = {https://doi.org/10.1111/j.1365-2966.2007.12519.x},
archivePrefix = {arXiv},
       eprint = {0709.4470},
 primaryClass = {astro-ph},
       adsurl = {https://ui.adsabs.harvard.edu/abs/2008MNRAS.383..525K}
}

@ARTICLE{Konar12,
       author = {{Konar}, C. and {Hardcastle}, M.~J. and {Jamrozy}, M. and {Croston}, J.~H. and {Nandi}, S.},
        title = "{Rejuvenated radio galaxies J0041+3224 and J1835+6204: how long can the quiescent phase of nuclear activity last?}",
      journal = {\mnras},
         year = 2012,
        
       volume = {424},
       number = {2},
        pages = {1061-1076},
          doi = {https://doi.org/10.1111/j.1365-2966.2012.21279.x},
       adsurl = {https://ui.adsabs.harvard.edu/abs/2012MNRAS.424.1061K}
}

@ARTICLE{Konar13b,
       author = {{Konar}, C. and {Hardcastle}, M.~J. and {Jamrozy}, M. and {Croston}, J.~H.},
        title = "{Episodic radio galaxies J0116-4722 and J1158+2621: can we constrain the quiescent phase of nuclear activity?}",
      journal = {\mnras},
         year = 2013,
        
       volume = {430},
       number = {3},
        pages = {2137-2153},
          doi = {https://doi.org/10.1093/mnras/stt040},
archivePrefix = {arXiv},
       eprint = {1309.1397},
 primaryClass = {astro-ph.CO},
       adsurl = {https://ui.adsabs.harvard.edu/abs/2013MNRAS.430.2137K}
}

@ARTICLE{Gabkrause11,
       author = {{Gaibler}, V. and {Khochfar}, S. and {Krause}, M.},
        title = "{Asymmetries in extragalactic double radio sources: clues from 3D simulations of jet-disc interaction}",
      journal = {\mnras},
         year = 2011,
        
       volume = {411},
       number = {1},
        pages = {155-161},
          doi = {https://doi.org/10.1111/j.1365-2966.2010.17674.x},
archivePrefix = {arXiv},
       eprint = {1008.2757},
 primaryClass = {astro-ph.HE},
       adsurl = {https://ui.adsabs.harvard.edu/abs/2011MNRAS.411..155G}
}

@ARTICLE{yatesjones21,
       author = {{Yates-Jones}, Patrick M. and {Shabala}, Stanislav S. and {Krause}, Martin G.~H.},
        title = "{Dynamics of relativistic radio jets in asymmetric environments}",
      journal = {\mnras},
         year = 2021,
        
       volume = {508},
       number = {4},
        pages = {5239-5250},
          doi = {https://doi.org/10.1093/mnras/stab2917},
archivePrefix = {arXiv},
       eprint = {2110.03162},
 primaryClass = {astro-ph.HE},
       adsurl = {https://ui.adsabs.harvard.edu/abs/2021MNRAS.508.5239Y}
}

@ARTICLE{4most,
       author = {{de Jong}, R.~S. and {Agertz}, O. and {Berbel}, A.~A. and {Aird}, J. and {Alexander}, D.~A. and {Amarsi}, A. and {Anders}, F. and {Andrae}, R. and {Ansarinejad}, B. and {Ansorge}, W. and {Antilogus}, P. and {Anwand-Heerwart}, H. and {Arentsen}, A. and {Arnadottir}, A. and {Asplund}, M. and {Auger}, M. and {Azais}, N. and {Baade}, D. and {Baker}, G. and {Baker}, S. and {Balbinot}, E. and {Baldry}, I.~K. and {Banerji}, M. and {Barden}, S. and {Barklem}, P. and {Barth{\'e}l{\'e}my-Mazot}, E. and {Battistini}, C. and {Bauer}, S. and {Bell}, C.~P.~M. and {Bellido-Tirado}, O. and {Bellstedt}, S. and {Belokurov}, V. and {Bensby}, T. and {Bergemann}, M. and {Bestenlehner}, J.~M. and {Bielby}, R. and {Bilicki}, M. and {Blake}, C. and {Bland-Hawthorn}, J. and {Boeche}, C. and {Boland}, W. and {Boller}, T. and {Bongard}, S. and {Bongiorno}, A. and {Bonifacio}, P. and {Boudon}, D. and {Brooks}, D. and {Brown}, M.~J.~I. and {Brown}, R. and {Br{\"u}ggen}, M. and {Brynnel}, J. and {Brzeski}, J. and {Buchert}, T. and {Buschkamp}, P. and {Caffau}, E. and {Caillier}, P. and {Carrick}, J. and {Casagrande}, L. and {Case}, S. and {Casey}, A. and {Cesarini}, I. and {Cescutti}, G. and {Chapuis}, D. and {Chiappini}, C. and {Childress}, M. and {Christlieb}, N. and {Church}, R. and {Cioni}, M.-R.~L. and {Cluver}, M. and {Colless}, M. and {Collett}, T. and {Comparat}, J. and {Cooper}, A. and {Couch}, W. and {Courbin}, F. and {Croom}, S. and {Croton}, D. and {Daguis{\'e}}, E. and {Dalton}, G. and {Davies}, L.~J.~M. and {Davis}, T. and {de Laverny}, P. and {Deason}, A. and {Dionies}, F. and {Disseau}, K. and {Doel}, P. and {D{\"o}scher}, D. and {Driver}, S.~P. and {Dwelly}, T. and {Eckert}, D. and {Edge}, A. and {Edvardsson}, B. and {Youssoufi}, D.~E. and {Elhaddad}, A. and {Enke}, H. and {Erfanianfar}, G. and {Farrell}, T. and {Fechner}, T. and {Feiz}, C. and {Feltzing}, S. and {Ferreras}, I. and {Feuerstein}, D. and {Feuillet}, D. and {Finoguenov}, A. and {Ford}, D. and {Fotopoulou}, S. and {Fouesneau}, M. and {Frenk}, C. and {Frey}, S. and {Gaessler}, W. and {Geier}, S. and {Gentile Fusillo}, N. and {Gerhard}, O. and {Giannantonio}, T. and {Giannone}, D. and {Gibson}, B. and {Gillingham}, P. and {Gonz{\'a}lez-Fern{\'a}ndez}, C. and {Gonzalez-Solares}, E. and {Gottloeber}, S. and {Gould}, A. and {Grebel}, E.~K. and {Gueguen}, A. and {Guiglion}, G. and {Haehnelt}, M. and {Hahn}, T. and {Hansen}, C.~J. and {Hartman}, H. and {Hauptner}, K. and {Hawkins}, K. and {Haynes}, D. and {Haynes}, R. and {Heiter}, U. and {Helmi}, A. and {Aguayo}, C.~H. and {Hewett}, P. and {Hinton}, S. and {Hobbs}, D. and {Hoenig}, S. and {Hofman}, D. and {Hook}, I. and {Hopgood}, J. and {Hopkins}, A. and {Hourihane}, A. and {Howes}, L. and {Howlett}, C. and {Huet}, T. and {Irwin}, M. and {Iwert}, O. and {Jablonka}, P. and {Jahn}, T. and {Jahnke}, K. and {Jarno}, A. and {Jin}, S. and {Jofre}, P. and {Johl}, D. and {Jones}, D. and {J{\"o}nsson}, H. and {Jordan}, C. and {Karovicova}, I. and {Khalatyan}, A. and {Kelz}, A. and {Kennicutt}, R. and {King}, D. and {Kitaura}, F. and {Klar}, J. and {Klauser}, U. and {Kneib}, J.-P. and {Koch}, A. and {Koposov}, S. and {Kordopatis}, G. and {Korn}, A. and {Kosmalski}, J. and {Kotak}, R. and {Kovalev}, M. and {Kreckel}, K. and {Kripak}, Y. and {Krumpe}, M. and {Kuijken}, K. and {Kunder}, A. and {Kushniruk}, I. and {Lam}, M.~I. and {Lamer}, G. and {Laurent}, F. and {Lawrence}, J. and {Lehmitz}, M. and {Lemasle}, B. and {Lewis}, J. and {Li}, B. and {Lidman}, C. and {Lind}, K. and {Liske}, J. and {Lizon}, J.-L. and {Loveday}, J. and {Ludwig}, H.-G. and {McDermid}, R.~M. and {Maguire}, K. and {Mainieri}, V. and {Mali}, S. and {Mandel}, H.},
        title = "{4MOST: Project overview and information for the First Call for Proposals}",
      journal = {The Messenger},
         year = 2019,
        
       volume = {175},
        pages = {3-11},
          doi = {https://doi.org/10.18727/0722-6691/5117},
archivePrefix = {arXiv},
       eprint = {1903.02464},
 primaryClass = {astro-ph.IM},
       adsurl = {https://ui.adsabs.harvard.edu/abs/2019Msngr.175....3D}
}

@ARTICLE{desi25,
       author = {{DESI Collaboration} and {Abdul-Karim}, M. and {Adame}, A.~G. and {Aguado}, D. and {Aguilar}, J. and {Ahlen}, S. and {Alam}, S. and {Aldering}, G. and {Alexander}, D.~M. and {Alfarsy}, R. and {Allen}, L. and {Allende Prieto}, C. and {Alves}, O. and {Anand}, A. and {Andrade}, U. and {Armengaud}, E. and {Avila}, S. and {Aviles}, A. and {Awan}, H. and {Bailey}, S. and {Baleato Lizancos}, A. and {Ballester}, O. and {Bault}, A. and {Bautista}, J. and {BenZvi}, S. and {Beraldo e Silva}, L. and {Bermejo-Climent}, J.~R. and {Beutler}, F. and {Bianchi}, D. and {Blake}, C. and {Blum}, R. and {Bolton}, A.~S. and {Bonici}, M. and {Brieden}, S. and {Brodzeller}, A. and {Brooks}, D. and {Buckley-Geer}, E. and {Burtin}, E. and {Canning}, R. and {Carnero Rosell}, A. and {Carr}, A. and {Carrilho}, P. and {Casas}, L. and {Castander}, F.~J. and {Cereskaite}, R. and {Cervantes-Cota}, J.~L. and {Chaussidon}, E. and {Chaves-Montero}, J. and {Chen}, S. and {Chen}, X. and {Claybaugh}, T. and {Cole}, S. and {Cooper}, A.~P. and {Cousinou}, M.-C. and {Cuceu}, A. and {Davis}, T.~M. and {Dawson}, K.~S. and {de Belsunce}, R. and {de la Cruz}, R. and {de la Macorra}, A. and {de Mattia}, A. and {Deiosso}, N. and {Della Costa}, J. and {Demina}, R. and {Demirbozan}, U. and {DeRose}, J. and {Dey}, A. and {Dey}, B. and {Ding}, J. and {Ding}, Z. and {Doel}, P. and {Douglass}, K. and {Dowicz}, M. and {Ebina}, H. and {Edelstein}, J. and {Eisenstein}, D.~J. and {Elbers}, W. and {Emas}, N. and {Escoffier}, S. and {Fagrelius}, P. and {Fan}, X. and {Fanning}, K. and {Fawcett}, V.~A. and {Fernandez-Garcia}, E. and {Ferraro}, S. and {Findlay}, N. and {Font-Ribera}, A. and {Forero-Romero}, J.~E. and {Forero-Sanchez}, D. and {Frenk}, C.~S. and {Gansicke}, B.~T. and {Galbany}, L. and {Garcia-Bellido}, J. and {Garcia-Quintero}, C. and {Garrison}, L.~H. and {Gaztanaga}, E. and {Gil-Marin}, H. and {Gnedin}, O.~Y. and {Gontcho}, S. Gontcho A and {Gonzalez-Morales}, A.~X. and {Gonzalez-Perez}, V. and {Gordon}, C. and {Graur}, O. and {Green}, D. and {Gruen}, D. and {Gsponer}, R. and {Guandalin}, C. and {Gutierrez}, G. and {Guy}, J. and {Hahn}, C. and {Han}, J.~J. and {Han}, J. and {He}, S. and {Herrera-Alcantar}, H.~K. and {Honscheid}, K. and {Hou}, J. and {Howlett}, C. and {Huterer}, D. and {Irsic}, V. and {Ishak}, M. and {Jacques}, A. and {Jimenez}, J. and {Jing}, Y.~P. and {Joachimi}, B. and {Joudaki}, S. and {Joyce}, R. and {Jullo}, E. and {Juneau}, S. and {Karacayli}, N.~G. and {Karim}, T. and {Kehoe}, R. and {Kent}, S. and {Khederlarian}, A. and {Kirkby}, D. and {Kisner}, T. and {Kitaura}, F.-S. and {Kizhuprakkat}, N. and {Kong}, H. and {Koposov}, S.~E. and {Kremin}, A. and {Krolewski}, A. and {Lahav}, O. and {Lai}, Y. and {Lamman}, C. and {Lan}, T.-W. and {Landriau}, M. and {Lang}, D. and {Lange}, J.~U. and {Lasker}, J. and {Le Goff}, J.~M. and {Le Guillou}, L. and {Leauthaud}, A. and {Levi}, M.~E. and {Li}, S. and {Li}, T.~S. and {Lodha}, K. and {Lokken}, M. and {Luo}, Y. and {Magneville}, C. and {Manera}, M. and {Manser}, C.~J. and {Margala}, D. and {Martini}, P. and {Maus}, M. and {McCullough}, J. and {McDonald}, P. and {Medina}, G.~E. and {Medina-Varela}, L. and {Meisner}, A. and {Mena-Fernandez}, J. and {Menegas}, A. and {Mezcua}, M. and {Miquel}, R. and {Montero-Camacho}, P. and {Moon}, J. and {Moustakas}, J. and {Munoz-Gutierrez}, A. and {Munoz-Santos}, D. and {Myers}, A.~D. and {Myles}, J. and {Nadathur}, S. and {Najita}, J. and {Napolitano}, L. and {Newman}, J.~A. and {Nikakhtar}, F. and {Nikutta}, R. and {Niz}, G. and {Noriega}, H.~E. and {Padmanabhan}, N. and {Paillas}, E. and {Palanque-Delabrouille}, N. and {Palmese}, A. and {Pan}, J. and {Pan}, Z. and {Parkinson}, D. and {Peacock}, J. and {Percival}, W.~J. and {Perez-Fernandez}, A. and {Perez-Rafols}, I. and {Peterson}, P.},
        title = "{Data Release 1 of the Dark Energy Spectroscopic Instrument}",
      journal = {arXiv e-prints},
         year = 2025,
          eid = {arXiv:2503.14745},
        pages = {arXiv:2503.14745},
          doi = {https://doi.org/10.48550/arXiv.2503.14745},
archivePrefix = {arXiv},
       eprint = {2503.14745},
 primaryClass = {astro-ph.CO},
       adsurl = {https://ui.adsabs.harvard.edu/abs/2025arXiv250314745D}
}

@ARTICLE{Shimwell17,
       author = {{Shimwell}, T.~W. and {R{\"o}ttgering}, H.~J.~A. and {Best}, P.~N. and {Williams}, W.~L. and {Dijkema}, T.~J. and {de Gasperin}, F. and {Hardcastle}, M.~J. and {Heald}, G.~H. and {Hoang}, D.~N. and {Horneffer}, A. and {Intema}, H. and {Mahony}, E.~K. and {Mandal}, S. and {Mechev}, A.~P. and {Morabito}, L. and {Oonk}, J.~B.~R. and {Rafferty}, D. and {Retana-Montenegro}, E. and {Sabater}, J. and {Tasse}, C. and {van Weeren}, R.~J. and {Br{\"u}ggen}, M. and {Brunetti}, G. and {Chy{\.z}y}, K.~T. and {Conway}, J.~E. and {Haverkorn}, M. and {Jackson}, N. and {Jarvis}, M.~J. and {McKean}, J.~P. and {Miley}, G.~K. and {Morganti}, R. and {White}, G.~J. and {Wise}, M.~W. and {van Bemmel}, I.~M. and {Beck}, R. and {Brienza}, M. and {Bonafede}, A. and {Calistro Rivera}, G. and {Cassano}, R. and {Clarke}, A.~O. and {Cseh}, D. and {Deller}, A. and {Drabent}, A. and {van Driel}, W. and {Engels}, D. and {Falcke}, H. and {Ferrari}, C. and {Fr{\"o}hlich}, S. and {Garrett}, M.~A. and {Harwood}, J.~J. and {Heesen}, V. and {Hoeft}, M. and {Horellou}, C. and {Israel}, F.~P. and {Kapi{\'n}ska}, A.~D. and {Kunert-Bajraszewska}, M. and {McKay}, D.~J. and {Mohan}, N.~R. and {Orr{\'u}}, E. and {Pizzo}, R.~F. and {Prandoni}, I. and {Schwarz}, D.~J. and {Shulevski}, A. and {Sipior}, M. and {Smith}, D.~J.~B. and {Sridhar}, S.~S. and {Steinmetz}, M. and {Stroe}, A. and {Varenius}, E. and {van der Werf}, P.~P. and {Zensus}, J.~A. and {Zwart}, J.~T.~L.},
        title = "{The LOFAR Two-metre Sky Survey. I. Survey description and preliminary data release}",
      journal = {\aap},
         year = 2017,
        
       volume = {598},
          eid = {A104},
        pages = {A104},
          doi = {https://doi.org/10.1051/0004-6361/201629313},
archivePrefix = {arXiv},
       eprint = {1611.02700},
 primaryClass = {astro-ph.IM},
       adsurl = {https://ui.adsabs.harvard.edu/abs/2017A&A...598A.104S}
}

@ARTICLE{Shimwell22,
       author = {{Shimwell}, T.~W. and {Hardcastle}, M.~J. and {Tasse}, C. and {Best}, P.~N. and {R{\"o}ttgering}, H.~J.~A. and {Williams}, W.~L. and {Botteon}, A. and {Drabent}, A. and {Mechev}, A. and {Shulevski}, A. and {van Weeren}, R.~J. and {Bester}, L. and {Br{\"u}ggen}, M. and {Brunetti}, G. and {Callingham}, J.~R. and {Chy{\.z}y}, K.~T. and {Conway}, J.~E. and {Dijkema}, T.~J. and {Duncan}, K. and {de Gasperin}, F. and {Hale}, C.~L. and {Haverkorn}, M. and {Hugo}, B. and {Jackson}, N. and {Mevius}, M. and {Miley}, G.~K. and {Morabito}, L.~K. and {Morganti}, R. and {Offringa}, A. and {Oonk}, J.~B.~R. and {Rafferty}, D. and {Sabater}, J. and {Smith}, D.~J.~B. and {Schwarz}, D.~J. and {Smirnov}, O. and {O'Sullivan}, S.~P. and {Vedantham}, H. and {White}, G.~J. and {Albert}, J.~G. and {Alegre}, L. and {Asabere}, B. and {Bacon}, D.~J. and {Bonafede}, A. and {Bonnassieux}, E. and {Brienza}, M. and {Bilicki}, M. and {Bonato}, M. and {Calistro Rivera}, G. and {Cassano}, R. and {Cochrane}, R. and {Croston}, J.~H. and {Cuciti}, V. and {Dallacasa}, D. and {Danezi}, A. and {Dettmar}, R.~J. and {Di Gennaro}, G. and {Edler}, H.~W. and {En{\ss}lin}, T.~A. and {Emig}, K.~L. and {Franzen}, T.~M.~O. and {Garc{\'\i}a-Vergara}, C. and {Grange}, Y.~G. and {G{\"u}rkan}, G. and {Hajduk}, M. and {Heald}, G. and {Heesen}, V. and {Hoang}, D.~N. and {Hoeft}, M. and {Horellou}, C. and {Iacobelli}, M. and {Jamrozy}, M. and {Jeli{\'c}}, V. and {Kondapally}, R. and {Kukreti}, P. and {Kunert-Bajraszewska}, M. and {Magliocchetti}, M. and {Mahatma}, V. and {Ma{\l}ek}, K. and {Mandal}, S. and {Massaro}, F. and {Meyer-Zhao}, Z. and {Mingo}, B. and {Mostert}, R.~I.~J. and {Nair}, D.~G. and {Nakoneczny}, S.~J. and {Nikiel-Wroczy{\'n}ski}, B. and {Orr{\'u}}, E. and {Pajdosz-{\'S}mierciak}, U. and {Pasini}, T. and {Prandoni}, I. and {van Piggelen}, H.~E. and {Rajpurohit}, K. and {Retana-Montenegro}, E. and {Riseley}, C.~J. and {Rowlinson}, A. and {Saxena}, A. and {Schrijvers}, C. and {Sweijen}, F. and {Siewert}, T.~M. and {Timmerman}, R. and {Vaccari}, M. and {Vink}, J. and {West}, J.~L. and {Wo{\l}owska}, A. and {Zhang}, X. and {Zheng}, J.},
        title = "{The LOFAR Two-metre Sky Survey. V. Second data release}",
      journal = {\aap},
         year = 2022,
        
       volume = {659},
          eid = {A1},
        pages = {A1},
          doi = {https://doi.org/10.1051/0004-6361/202142484},
archivePrefix = {arXiv},
       eprint = {2202.11733},
 primaryClass = {astro-ph.GA},
       adsurl = {https://ui.adsabs.harvard.edu/abs/2022A&A...659A...1S}
}

@INPROCEEDINGS{MIGHTEE,
       author = {{Jarvis}, M. and {Taylor}, R. and {Agudo}, I. and {Allison}, J.~R. and {Deane}, R.~P. and {Frank}, B. and {Gupta}, N. and {Heywood}, I. and {Maddox}, N. and {McAlpine}, K. and {Santos}, M. and {Scaife}, A.~M.~M. and {Vaccari}, M. and {Zwart}, J.~T.~L. and {Adams}, E. and {Bacon}, D.~J. and {Baker}, A.~J. and {Bassett}, B.~A. and {Best}, P.~N. and {Beswick}, R. and {Blyth}, S. and {Brown}, M.~L. and {Bruggen}, M. and {Cluver}, M. and {Colafrancesco}, S. and {Cotter}, G. and {Cress}, C. and {Dav{\'e}}, R. and {Ferrari}, C. and {Hardcastle}, M.~J. and {Hale}, C.~L. and {Harrison}, I. and {Hatfield}, P.~W. and {Klockner}, H.~R. and {Kolwa}, S. and {Malefahlo}, E. and {Marubini}, T. and {Mauch}, T. and {Moodley}, K. and {Morganti}, R. and {Norris}, R.~P. and {Peters}, J.~A. and {Prandoni}, I. and {Prescott}, M. and {Oliver}, S. and {Oozeer}, N. and {Rottgering}, H.~J.~A. and {Seymour}, N. and {Simpson}, C. and {Smirnov}, O. and {Smith}, D.~J.~B.},
        title = "{The MeerKAT International GHz Tiered Extragalactic Exploration (MIGHTEE) Survey}",
    booktitle = {MeerKAT Science: On the Pathway to the SKA},
         year = 2016,
        
          eid = {6},
        pages = {6},
          doi = {https://doi.org/10.22323/1.277.0006},
archivePrefix = {arXiv},
       eprint = {1709.01901},
 primaryClass = {astro-ph.GA},
       adsurl = {https://ui.adsabs.harvard.edu/abs/2016mks..confE...6J}
}

@ARTICLE{ceglowski13,
       author = {{Ceg{\l}owski}, M. and {Gawro{\'n}ski}, M.~P. and {Kunert-Bajraszewska}, M.},
        title = "{Orientation of the cores of hybrid morphology radio sources}",
      journal = {\aap},
         year = 2013,
        
       volume = {557},
          eid = {A75},
        pages = {A75},
          doi = {https://doi.org/10.1051/0004-6361/201220544},
archivePrefix = {arXiv},
       eprint = {1307.8255},
 primaryClass = {astro-ph.CO},
       adsurl = {https://ui.adsabs.harvard.edu/abs/2013A&A...557A..75C}
}

@ARTICLE{GK04,
       author = {{Gopal-Krishna} and {Wiita}, Paul J.},
        title = "{Asymmetries in Powerful Extragalactic Radio Sources}",
      journal = {arXiv e-prints},
         year = 2004,
        
          eid = {astro-ph/0409761},
        pages = {astro-ph/0409761},
          doi = {https://doi.org/10.48550/arXiv.astro-ph/0409761},
archivePrefix = {arXiv},
       eprint = {astro-ph/0409761},
 primaryClass = {astro-ph},
       adsurl = {https://ui.adsabs.harvard.edu/abs/2004astro.ph..9761G}
}

@ARTICLE{Edwards2010,
       author = {{Edwards}, Louise O.~V. and {Fadda}, Dario and {Frayer}, David T.},
        title = "{The First Bent Double Lobe Radio Source in a Known Cluster Filament: Constraints on the Intrafilament Medium}",
      journal = {\apjl},
         year = 2010,
        
       volume = {724},
       number = {2},
        pages = {L143-L147},
          doi = {https://doi.org/10.1088/2041-8205/724/2/L143},
archivePrefix = {arXiv},
       eprint = {1010.5289},
 primaryClass = {astro-ph.CO},
       adsurl = {https://ui.adsabs.harvard.edu/abs/2010ApJ...724L.143E}
}

@ARTICLE{Hota2022,
       author = {{Hota}, Ananda and {Dabhade}, Pratik and {Vaddi}, Sravani and {Konar}, Chiranjib and {Pal}, Sabyasachi and {Gulati}, Mamta and {Stalin}, C.~S. and {Avinash}, Ck and {Kumar}, Avinash and {Rajoria}, Megha and {Purohit}, Arundhati},
        title = "{RAD@home citizen science discovery of an active galactic nucleus spewing a large unipolar radio bubble on to its merging companion galaxy}",
      journal = {\mnras},
         year = 2022,
        
       volume = {517},
       number = {1},
        pages = {L86-L91},
          doi = {https://doi.org/10.1093/mnrasl/slac116},
archivePrefix = {arXiv},
       eprint = {2210.06100},
 primaryClass = {astro-ph.GA},
       adsurl = {https://ui.adsabs.harvard.edu/abs/2022MNRAS.517L..86H}
}

@ARTICLE{Mahato2025,
       author = {{Mahato}, Mousumi and {Tempel}, Elmo and {Sankhyayan}, Shishir and {Dabhade}, Pratik and {Chavan}, Kshitij},
        title = "{Search and analysis of giant radio galaxies with associated nuclei (SAGAN): VI: When jets meet filaments ─ Environmental imprints on the growth of giant radio galaxies}",
      journal = {\aap},
         year = 2026,
        
       volume = {706},
          eid = {A310},
        pages = {A310},
          doi = {https://doi.org/10.1051/0004-6361/202557646},
archivePrefix = {arXiv},
       eprint = {2512.07985},
 primaryClass = {astro-ph.GA},
       adsurl = {https://ui.adsabs.harvard.edu/abs/2026A&A...706A.310M}
}

@ARTICLE{Bagchi2007,
       author = {{Bagchi}, Joydeep and {Gopal-Krishna} and {Krause}, Marita and {Joshi}, Santosh},
        title = "{A Giant Radio Jet Ejected by an Ultramassive Black Hole in a Single-lobed Radio Galaxy}",
      journal = {\apjl},
         year = 2007,
        
       volume = {670},
       number = {2},
        pages = {L85-L88},
          doi = {https://doi.org/10.1086/524220},
archivePrefix = {arXiv},
       eprint = {0712.0543},
 primaryClass = {astro-ph},
       adsurl = {https://ui.adsabs.harvard.edu/abs/2007ApJ...670L..85B}
}

@ARTICLE{Harris1984,
       author = {{Harris}, D.~E. and {Costain}, C.~H. and {Dewdney}, P.~E.},
        title = "{The X-ray features of a single-lobed radio galaxy in Abell 754.}",
      journal = {\apj},
         year = 1984,
        
       volume = {280},
        pages = {532-541},
          doi = {https://doi.org/10.1086/162022},
       adsurl = {https://ui.adsabs.harvard.edu/abs/1984ApJ...280..532H}
}

@ARTICLE{Croft2006,
       author = {{Croft}, Steve and {van Breugel}, Wil and {de Vries}, Wim and {Dopita}, Mike and {Martin}, Chris and {Morganti}, Raffaella and {Neff}, Susan and {Oosterloo}, Tom and {Schiminovich}, David and {Stanford}, S.~A. and {van Gorkom}, Jacqueline},
        title = "{Minkowski's Object: A Starburst Triggered by a Radio Jet, Revisited}",
      journal = {\apj},
         year = 2006,
        
       volume = {647},
       number = {2},
        pages = {1040-1055},
          doi = {https://doi.org/10.1086/505526},
archivePrefix = {arXiv},
       eprint = {astro-ph/0604557},
 primaryClass = {astro-ph},
       adsurl = {https://ui.adsabs.harvard.edu/abs/2006ApJ...647.1040C}
}

@ARTICLE{Best95,
       author = {{Best}, P.~N. and {Bailer}, D.~M. and {Longair}, M.~S. and {Riley}, J.~M.},
        title = "{Radio source asymmetries and unified schemes}",
      journal = {\mnras},
         year = 1995,
        
       volume = {275},
       number = {4},
        pages = {1171-1184},
          doi = {https://doi.org/10.1093/mnras/275.4.1171},
       adsurl = {https://ui.adsabs.harvard.edu/abs/1995MNRAS.275.1171B}
}

@ARTICLE{Garrington1988,
       author = {{Garrington}, S.~T. and {Leahy}, J.~P. and {Conway}, R.~G. and {Laing}, R.~A.},
        title = "{A systematic asymmetry in the polarization properties of double radio sources with one jet}",
      journal = {\nat},
         year = 1988,
        
       volume = {331},
       number = {6152},
        pages = {147-149},
          doi = {https://doi.org/10.1038/331147a0},
       adsurl = {https://ui.adsabs.harvard.edu/abs/1988Natur.331..147G}
}

@ARTICLE{Garrington1991,
       author = {{Garrington}, S.~T. and {Conway}, R.~G. and {Leahy}, J.~P.},
        title = "{Asymmetry depolarization in double radio sources with one sided jets.}",
      journal = {\mnras},
         year = 1991,
        
       volume = {250},
        pages = {171},
          doi = {https://doi.org/10.1093/mnras/250.1.171},
       adsurl = {https://ui.adsabs.harvard.edu/abs/1991MNRAS.250..171G}
}

@ARTICLE{Laing1988,
       author = {{Laing}, R.~A.},
        title = "{The sidedness of jets and depolarization in powerful extragalactic radio sources}",
      journal = {\nat},
         year = 1988,
        
       volume = {331},
       number = {6152},
        pages = {149-151},
          doi = {https://doi.org/10.1038/331149a0},
       adsurl = {https://ui.adsabs.harvard.edu/abs/1988Natur.331..149L}
}

@ARTICLE{Kumari2024,
       author = {{Kumari}, Shobha and {Pal}, Sabyasachi and {Hardcastle}, Martin J. and {Horton}, Maya A.},
        title = "{J0011+3217: A peculiar radio galaxy with a one-sided secondary lobe and misaligned giant primary lobes}",
      journal = {\aap},
         year = 2024,
        
       volume = {689},
          eid = {A301},
        pages = {A301},
          doi = {https://doi.org/10.1051/0004-6361/202347367},
archivePrefix = {arXiv},
       eprint = {2406.14889},
 primaryClass = {astro-ph.GA},
       adsurl = {https://ui.adsabs.harvard.edu/abs/2024A&A...689A.301K}
}

@ARTICLE{Nandi2010,
  title = {A multifrequency study of the large radio galaxies 3C46 and 3C452},
  ISSN = {1365-2966},
  DOI = {https://doi.org/10.1111/j.1365-2966.2010.16286.x},
  journal = {Monthly Notices of the Royal Astronomical Society},
  publisher = {Oxford University Press (OUP)},
  author = {Nandi,  S. and Pirya,  A. and Pal,  S. and Konar,  C. and Saikia,  D. J. and Singh,  M.},
  year = {2010}
}

@article{Patra2024,
  title = {Multi-frequency study of large size radio galaxies 3C 35 and 3C 284},
  volume = {73},
  ISSN = {0273-1177},
  DOI = {https://doi.org/10.1016/j.asr.2023.07.070},
  number = {1},
  journal = {Advances in Space Research},
  publisher = {Elsevier BV},
  author = {Patra, Dusmanta and Pal, Sabyasachi},
  year = {2024},
  
  pages = {1113–1124}
}

@ARTICLE{Shimwell2026,
       author = {{Shimwell}, T.~W. and {Hardcastle}, M.~J. and {Tasse}, C. and {Drabent}, A. and {Botteon}, A. and {Williams}, W.~L. and {Best}, P.~N. and {R{\"o}ttgering}, H.~J.~A. and {Br{\"u}ggen}, M. and {Brunetti}, G. and {Callingham}, J.~R. and {Chy{\.z}y}, K.~T. and {Conway}, J.~E. and {De Gasperin}, F. and {Haverkorn}, M. and {Horellou}, C. and {Jackson}, N. and {Miley}, G.~K. and {Morabito}, L.~K. and {Morganti}, R. and {O'Sullivan}, S.~P. and {Schwarz}, D.~J. and {Smith}, D.~J.~B. and {van Weeren}, R.~J. and {Vedantham}, H.~K. and {White}, G.~J. and {Ahmadi}, A. and {Alegre}, L. and {Arias}, M. and {Asabere}, B. and {Bahr-Kalus}, B. and {Barkus}, B. and {Bilicki}, M. and {B{\"o}hme}, L. and {Brentjens}, M. and {Brienza}, M. and {Bomans}, D.~J. and {Bonafede}, A. and {Bonato}, M. and {Bonnassieux}, E. and {Boxelaar}, J.~M. and {Camera}, S. and {Cassano}, R. and {Chilufya}, J. and {Cianfaglione}, M. and {Croston}, J.~H. and {Cuciti}, V. and {Dabhade}, P. and {De Rubeis}, E. and {de Jong}, J.~M.~G.~H.~J. and {Dallacasa}, D. and {Dettmar}, R.~J. and {Duncan}, K.~J. and {Di Gennaro}, G. and {Edler}, H.~W. and {Groeneveld}, C. and {G{\"u}rkan}, G. and {Hajduk}, M. and {Hale}, C.~L. and {Heesen}, V. and {Hoang}, D.~N. and {Hoeft}, M. and {Holties}, H. and {Horton}, M.~A. and {Iacobelli}, M. and {Jamrozy}, M. and {Jarvis}, M.~J. and {Jelic}, V. and {Kadler}, M. and {Kondapally}, R. and {Kunert-Bajraszewska}, M. and {Loose}, M. and {Magliocchetti}, M. and {Ma{\l}ek}, K. and {Manzano}, C. and {McKean}, J.~P. and {Mevius}, M. and {Mingo}, B. and {Miskolczi}, A. and {Misra}, A. and {Mold{\'o}n}, J. and {Nair}, D.~G. and {Nakoneczny}, S.~J. and {Orru}, E. and {Pashapour-Ahmadabadi}, M. and {Pasini}, T. and {Petley}, J. and {Pierce}, J.~C.~S. and {Prandoni}, I. and {Rafferty}, D. and {Rajpurohit}, K. and {Riseley}, C.~J. and {Roberts}, I.~D. and {Sethi}, S. and {Shulevski}, A. and {Stein}, M. and {Stuardi}, C. and {Sweijen}, F. and {ter Veen}, S. and {Timmerman}, R. and {Vaccari}, M. and {Wijnholds}, S.},
        title = "{The LOFAR Two-metre Sky Survey: VII. Third Data Release}",
      journal = {\aap},
         year = 2026,
        
       volume = {707},
          eid = {A198},
        pages = {A198},
          doi = {https://doi.org/10.1051/0004-6361/202557749},
archivePrefix = {arXiv},
       eprint = {2602.15949},
 primaryClass = {astro-ph.GA},
       adsurl = {https://ui.adsabs.harvard.edu/abs/2026A&A...707A.198S}
}

@ARTICLE{Hale2025,
       author = {{Hale}, C.~L. and {Heywood}, I. and {Jarvis}, M.~J. and {Whittam}, I.~H. and {Best}, P.~N. and {An}, Fangxia and {Bowler}, R.~A.~A. and {Harrison}, I. and {Matthews}, A. and {Smith}, D.~J.~B. and {Taylor}, A.~R. and {Vaccari}, M.},
        title = "{MIGHTEE: the continuum survey Data Release 1}",
      journal = {\mnras},
         year = 2025,
        
       volume = {536},
       number = {3},
        pages = {2187-2211},
          doi = {https://doi.org/10.1093/mnras/stae2528},
archivePrefix = {arXiv},
       eprint = {2411.04958},
 primaryClass = {astro-ph.GA},
       adsurl = {https://ui.adsabs.harvard.edu/abs/2025MNRAS.536.2187H}
}

@ARTICLE{Norris2021,
       author = {{Norris}, Ray P. and {Marvil}, Joshua and {Collier}, J.~D. and {Kapi{\'n}ska}, Anna D. and {O'Brien}, Andrew N. and {Rudnick}, L. and {Andernach}, Heinz and {Asorey}, Jacobo and {Brown}, Michael J.~I. and {Br{\"u}ggen}, Marcus and {Crawford}, Evan and {English}, Jayanne and {Rahman}, Syed Faisal ur and {Filipovi{\'c}}, Miroslav D. and {Gordon}, Yjan and {G{\"u}rkan}, G{\"u}lay and {Hale}, Catherine and {Hopkins}, Andrew M. and {Huynh}, Minh T. and {HyeongHan}, Kim and {James Jee}, M. and {Koribalski}, B{\"a}rbel S. and {Lenc}, Emil and {Luken}, Kieran and {Parkinson}, David and {Prandoni}, Isabella and {Raja}, Wasim and {Reiprich}, Thomas H. and {Riseley}, Christopher J. and {Shabala}, Stanislav S. and {Sheil}, Jaimie R. and {Vernstrom}, Tessa and {Whiting}, Matthew T. and {Allison}, James R. and {Anderson}, C.~S. and {Ball}, Lewis and {Bell}, Martin and {Bunton}, John and {Galvin}, T.~J. and {Gupta}, Neeraj and {Hotan}, Aidan and {Jacka}, Colin and {Macgregor}, Peter J. and {Mahony}, Elizabeth K. and {Maio}, Umberto and {Moss}, Vanessa and {Pandey-Pommier}, M. and {Voronkov}, Maxim A.},
        title = "{The Evolutionary Map of the Universe pilot survey}",
      journal = {\pasa},
         year = 2021,
        
       volume = {38},
          eid = {e046},
        pages = {e046},
          doi = {https://doi.org/10.1017/pasa.2021.42},
archivePrefix = {arXiv},
       eprint = {2108.00569},
 primaryClass = {astro-ph.CO},
       adsurl = {https://ui.adsabs.harvard.edu/abs/2021PASA...38...46N}
}

@ARTICLE{Bonaldi2024,
  author       = {Bonaldi, A. and Zwaan, M. and Best, P. and Broderick, J. 
                  and Goedhart, S. and Mahony, E.},
  title        = {{ESO--SKAO} Synergies: {SKAO}, {SKA} Precursors/Pathfinders 
                  and {ESO} Facilities},
  journal      = {The Messenger},
  year         = {2024},
  volume       = {193},
  pages        = {5--8},
  doi          = {https://doi.org/10.18727/0722-6691/5361}
}

@techreport{Braun2024,
  author       = {Braun, R. and {others}},
  title        = {Anticipated Science Performance of the {SKA Observatory} 
                  Deployment Baseline Telescopes},
  institution  = {SKA Observatory},
  year         = {2024},
  type         = {SKAO Technical Document},
  number       = {SKA-TEL-SKO-0000818},
  note         = {Revision undergoing update; see also Braun et al.\ 2019 
                  for the earlier version}
}

@ARTICLE{deJong2024,
       author = {{de Jong}, J.~M.~G.~H.~J. and {van Weeren}, R.~J. and {Sweijen}, F. and {Oonk}, J.~B.~R. and {Shimwell}, T.~W. and {Offringa}, A.~R. and {Morabito}, L.~K. and {R{\"o}ttgering}, H.~J.~A. and {Kondapally}, R. and {Escott}, E.~L. and {Best}, P.~N. and {Bondi}, M. and {Ye}, H. and {Petley}, J.~W.},
        title = "{Into the depths: Unveiling ELAIS-N1 with LOFAR's deepest sub-arcsecond wide-field images}",
      journal = {\aap},
         year = 2024,
        
       volume = {689},
          eid = {A80},
        pages = {A80},
          doi = {https://doi.org/10.1051/0004-6361/202450595},
archivePrefix = {arXiv},
       eprint = {2407.13247},
 primaryClass = {astro-ph.IM},
       adsurl = {https://ui.adsabs.harvard.edu/abs/2024A&A...689A..80D}
}

@ARTICLE{Shimwell2025,
       author = {{Shimwell}, T.~W. and {Hale}, C.~L. and {Best}, P.~N. and {Botteon}, A. and {Drabent}, A. and {Hardcastle}, M.~J. and {Jeli{\'c}}, V. and {de Jong}, J.~M.~G.~H.~J. and {Kondapally}, R. and {R{\"o}ttgering}, H.~J.~A. and {Tasse}, C. and {van Weeren}, R.~J. and {Williams}, W.~L. and {Bonafede}, A. and {Bondi}, M. and {Br{\"u}ggen}, M. and {Brunetti}, G. and {Callingham}, J.~R. and {De Gasperin}, F. and {Duncan}, K.~J. and {Horellou}, C. and {Iyer}, S. and {de Ruiter}, I. and {Ma{\l}ek}, K. and {Nair}, D.~G. and {Morabito}, L.~K. and {Prandoni}, I. and {Rowlinson}, A. and {Sabater}, J. and {Shulevski}, A. and {Smith}, D.~J.~B. and {Sweijen}, F.},
        title = "{The LOFAR Two-metre Sky Survey: Deep Fields Data Release 2: I. The ELAIS-N1 field}",
      journal = {\aap},
         year = 2025,
        
       volume = {695},
          eid = {A80},
        pages = {A80},
          doi = {https://doi.org/10.1051/0004-6361/202452930},
archivePrefix = {arXiv},
       eprint = {2501.04093},
 primaryClass = {astro-ph.CO},
       adsurl = {https://ui.adsabs.harvard.edu/abs/2025A&A...695A..80S}
}

@ARTICLE{Hota2026,
       author = {{Hota}, Ananda and {Dabhade}, Pratik and {Ghosh}, Shubhrangshu and {Limbo}, Pranim and {Konar}, C. and {Sethi}, Sagar and {Manik}, Souvik and {Sahasranshu}, Aditya and {Pal}, Sabyasachi and {Damle}, Mitali and {Vaddi}, Sravani and {Purohit}, Arundhati},
        title = "{RAD@home discovery of a bow-and-arrow radio galaxy tracing a {\ensuremath{\sim}}560 kpc bow-shock structure in a multihalo environment}",
      journal = {\mnras},
         year = 2026,
        month = jul,
       volume = {549},
       number = {4},
          eid = {stag1033},
        pages = {stag1033},
          doi = {10.1093/mnras/stag1033},
       adsurl = {https://ui.adsabs.harvard.edu/abs/2026MNRAS.549g1033H}
}

\end{document}